\documentclass{jfm}

\usepackage{graphicx,epsfig,subfig}
\usepackage{epstopdf}
\usepackage{natbib}
\usepackage{amsmath}
\usepackage{mathtools}
\usepackage{amssymb}
\usepackage{color}
\usepackage{setspace}
\usepackage{mathrsfs}
\usepackage[usenames,dvipsnames]{xcolor}
\usepackage{pstool}
\usepackage[linesnumbered,ruled]{algorithm2e}

\newcommand\dd{\mathrm{d}}
\newcommand\zi{\mathrm{i}}

\def\bdot{\raise.2em\hbox to .15em{.}}

\definecolor{gray}{gray}{0.8}
\definecolor{darkgray}{gray}{0.5}

\def\bdotblack{\raise.25em\hbox to .15em{.}}

\title[Koopman expansions]{Koopman mode expansions between simple invariant solutions}

\author[J. Page \& R. R. Kerswell]%
{Jacob Page \& Rich R. Kerswell} 

\affiliation{DAMTP, Center for Mathematical Sciences, University of Cambridge, Cambridge, CB3 0WA, UK}

\pubyear{2018}
\volume{}
\pagerange{}
\date{\today}
\begin{document}

\maketitle

\begin{abstract}

A Koopman decomposition is a powerful method of analysis for fluid flows leading to an apparently linear description of nonlinear dynamics in which the flow is expressed as a superposition of fixed spatial structures with exponential time dependence.  Attempting a Koopman decomposition is simple in practice due to a connection with Dynamic Mode Decomposition (DMD).    However, there are non-trivial requirements for the Koopman decomposition and DMD to overlap which mean it is often difficult to establish whether the latter is truly approximating the former. Here, we focus on nonlinear systems containing multiple simple invariant solutions where it is unclear how to construct a consistent Koopman decomposition, or how DMD might be applied to locate these solutions. First, we derive a Koopman decomposition for a heteroclinic connection in a Stuart-Landau equation revealing two possible expansions. The expansions are centred about the two fixed points of the equation and extend beyond their linear subspaces before breaking down at a crossover point in state space. Well-designed DMD can extract the two expansions provided that the time window does not contain this crossover point.    We then apply DMD to the Navier-Stokes equations near to a heteroclinic connection in low-Reynolds number ($Re=O(100)$) plane Couette flow where there are multiple simple invariant solutions beyond the constant shear basic state.  This reveals as many different Koopman decompositions as simple invariant solutions present and again indicates the existence of crossover points between the expansions in state space.  Again,   DMD can extract these expansions only if it does not include a crossover point.
\end{abstract}


\section{Introduction}

In the past few decades the discovery of non-trivial exact solutions of the Navier-Stokes equations has given rise to a dynamical systems view of turbulent flow \citep{Kerswell2005, Eckhardt2007, Kawahara2012}. In this perspective, a turbulent orbit wanders in phase space between these so-called \emph{exact coherent structures} or simple invariant solutions  (e.g. equilibria, travelling waves, periodic orbits), pulled in along their stable manifolds and thrown out along their unstable manifolds \citep{Gibson2008,Gibson2009}.  Individually, exact coherent structures can offer a useful perspective on the fully turbulent dynamics:
their averaged properties often share qualitative similarities to statistics of the turbulence while their simple time dependence makes the underlying physical mechanisms far simpler to extract and analyse  \citep[e.g.][]{Waleffe1997,Kawahara2001, Wang2007,Hall2010}. Since the discovery of the first pair of non-trivial equilibria in plane Couette flow by \citet{Nagata1990}, exact coherent structures have been found in a wide range of flow geometries \citep{Waleffe1997, Waleffe2001, Faisst2003, Wedin2004, Gibson2008, Gibson2009, Uhlmann2010}, in spatially extended flows \citep{Schneider2010, Avila2013, Chantry2014,Zammert2014, GibsonBrand2014, BrandGibson2014}  and in stratified fluids \citep{Olvera2017,Deguchi2017,Lucas2017}.

A crucial step in attempting to converge exact solutions of the Navier-Stokes equations is the generation of an initial guess for the structure of interest, which is then fed into a Newton-Raphson algorithm. 
For an equilibrium the guess takes the the form of a velocity snapshot; for periodic orbits the snapshot must be supplemented with a guess for the period.  Currently, methods for generating guesses include (i) edge tracking \citep{Schneider2008}, (ii) using snapshots of turbulence \citep[equilibria only,][]{Gibson2009}, (iii) branch continuation of known solutions \citep{Nagata1990, Waleffe2001, Faisst2003, Wedin2004} or (iv) a recurrent flow analysis \citep{Kawahara2001,Viswanath2007,Cvitanovic2010, Chandler2013}. Each of these approaches has weaknesses, for example continuation cannot find unconnected solutions while recurrent flow analysis requires the turbulent flow to shadow a periodic orbit for at least one cycle -- increasingly improbable as the Reynolds number is increased.  

The recent emergence of \emph{Dynamic Mode Decomposition} (DMD) suggests an alternative approach to finding exact coherent structures in nonlinear simulation data.
DMD was originally invented by \citet{Schmid2010} as a post-processing technique for simulation or experimental data, with many variants on the algorithm developed since \citep[e.g.][]{Jovanovic2014,Williams2015}. DMD finds a linear operator that best maps (in a least squares sense) between equispaced snapshots of the flow. As a result of which the flow can then be expressed as a superposition of dynamic ``modes'' (eigenvectors of the DMD operator) with an exponential dependence on time.  An attractive feature of the method is that it can identify frequencies of oscillation in the flow which correspond to periods far longer than the time window over which observations are recorded. Beyond fluid mechanics, DMD has already been applied in areas as diverse as video processing \citep{Kutz2016_vid} and neuroscience \citep{BruntonB2016}.

Connecting the output of DMD with exact coherent structures rests on its connection to the Koopman operator which is a linear infinite dimensional operator that evolves functionals (or \emph{observables}) of the velocity field forward in time \citep{Koopman1931,Mezic2005, Mezic2013}.  The hope in Koopman operator theory \citep[proveable in some situations --][]{Mezic2005} is that the nonlinear evolution of any observable of the state $\mathbf u$ can be expressed as a sum of fixed spatial structures (Koopman modes) with an exponential time dependence. This is accomplished through a projection onto eigenfunctions of the Koopman operator, which are special scalar observables of the system which evolve like $\text{exp}(\lambda t)$, where $\lambda$ is the associated Koopman eigenvalue. 
Neutral Koopman eigenvalues can coincide with equilibria of the system whereas purely imaginary eigenvalues can identify harmonics of periodic orbits \citep{Mezic2005,Mezic2013}.  

Koopman eigenfunctions have been obtained analytically in some simple nonlinear ordinary differential equations \citep[e.g.][]{Bagheri2013,Brunton2016,Rowley2017} and recently for Burgers' equation which can be linearized by the Cole-Hopf transformation \citep{Page2018}. However, it is unlikely that closed-form expressions for Koopman eigenfunctions of the Navier-Stokes equations can be written down.
While there have been some ingenious attempts to discover Koopman eigenfunctions from nonlinear data \citep{Lusch2016}, these have so far been restricted to low-dimensional examples.
Instead, most studies focus on extracting the Koopman modes \citep{Rowley2009,Bagheri2013,Tu2014}, which under certain requirements overlap with dynamic modes obtained in DMD \citep{Tu2014,Williams2015}. There are two requirements for DMD and Koopman to coincide: (i) that sufficient data is available and (ii) that the Koopman eigenfunctions can be expressed as a linear combination of the functionals of the state which serve as the inputs to the DMD algorithm \citep{Williams2015}. 
The second point is difficult to enforce in practice and various strategies have been proposed to ensure the input function space is sufficiently `rich' \citep[e.g. `Kernel' based methods, see][]{DMDkutz}. 
Furthermore, even if DMD can accurately extract Koopman eigenfunctions, there is no guarantee that these then form a basis for the state variable itself \citep[e.g][]{Brunton2016,Page2018}. 
Alongside DMD, other related methods have been proposed to extract Koopman modes from turbulent flows that may circumvent some of these issues. For example, \citet{Arbabi2017} proposed an approach based on harmonic averaging to extract Koopman modes in high-Reynolds number  lid-driven cavity flow, while \citet{Sharma2016} demonstrated a connection between Koopman modes and modes of the resolvent operator.  

Our focus in this study is on the utility of DMD as a tool to identify exact coherent structures and their stable and unstable manifolds.  In systems with more than one simple invariant solution there are known issues related to both DMD and Koopman expansions. For example, \citet{Brunton2016} have demonstrated that it is not possible to form a Koopman invariant subspace that contains the state variable itself in a nonlinear system with more than one fixed point, indicating that there is not a single uniformly valid Koopman expansion. This fact may have implications for DMD and its ability to find Koopman eigenvalues, and there is reason to believe that this issue has been encountered in past studies. \citet{Bagheri2013} performed a multiple-scales analysis to analytically construct a Koopman decomposition for flow past a cylinder just beyond the critical Reynolds number. The expansion describes the transient collapse onto the oscillatory limit cycle (vortex shedding). However, \citet{Bagheri2013} could only match his analytical result to the output of DMD provided that the DMD observation window did not stretch too far back into the region of ``transient amplification''.  \citet{Eaves2016} found a similar results when performing DMD on a flow trajectory approaching and then receding from the fixed point edge state in small-box plane Couette flow.

In this paper we seek to bring some clarity to these issues by considering a pair of examples: a model ODE system with two fixed points and the Navier-Stokes equations with multiple solutions. We demonstrate that each simple invariant solution has an associated Koopman expansion for the state variable which extends beyond the respective linear subspace but which  breaks down at a point in state space. These \emph{crossover points} impact the ability of DMD to extract a Koopman decomposition from the data -- a DMD calculation with an observation window including a crossover point will fail. The structure of the remainder of this paper is as follows. In \S\ref{sec:SL} we briefly review the basics of Koopman mode decompositions before describing a new approach for their computation given a solution to a nonlinear equation. The results are applied to the Stuart-Landau equation and compared to DMD. In \S\ref{sec:HC} we extend these ideas to the Navier-Stokes equations, using DMD to find Koopman mode decompositions along heteroclinic connections between simple invariant solutions in low Reynolds plane Couette flow. Finally, concluding remarks are provided in \S\ref{sec:conc}. 


\section{Koopman mode decompositions of a Stuart-Landau equation}
\label{sec:SL}
%

\subsection{The Koopman operator}
We consider nonlinear dynamical systems of the form
\begin{equation}
    \partial_t \mathbf u = \mathbf F(\mathbf u),
    \label{eqn:nonlinear_eqn}
\end{equation}
with the time-forward map $\mathbf f^t(\mathbf u) = \mathbf u + \int_0^t \mathbf F(\mathbf u) \dd t'$.
The Koopman operator, $\mathscr{K}^t$, is an infinite-dimensional linear operator that propagates functionals $\psi$ of the state vector -  or \emph{``observables''} - forward in time \citep{Koopman1931,Mezic2005} along a trajectory of (\ref{eqn:nonlinear_eqn}),
\begin{equation}
    \mathscr{K}^t \psi(\mathbf u) := \psi(\mathbf f^t(\mathbf u)).
\end{equation}
The eigenfunctions of this linear, infinite-dimensional operator are special observables with exponential time dependence 
\begin{equation}
    \mathscr K^t \varphi_{\lambda}(\mathbf u) = \varphi_{\lambda}(\mathbf u)e^{\lambda t}.
\end{equation}
It then follows that  Koopman eigenfunctions can be computed by the relation
\begin{equation}
 \partial_t\varphi_{\lambda}(\mathbf u) =  \mathbf F(\mathbf u)  \cdot   \bnabla_{\mathbf u}\varphi_{\lambda}(\mathbf u)= \lambda \varphi_{\lambda}(\mathbf u).
    \label{eqn:koop_ef}
\end{equation}
The default assumption is then that the Koopman eigenfunctions can be used to expand a vector of observables,
\begin{equation}
    \boldsymbol \psi(\mathbf u) = \sum_n \varphi_{\lambda_n}(\mathbf u)\hat{\boldsymbol \psi}_n,
    \label{eqn:kmd}
\end{equation}
where the coefficients $\hat{\boldsymbol \psi}_n$ are called Koopman modes \citep{Rowley2009}.  A common choice is to consider a spatially-varying functional  so that the vector $ \boldsymbol \psi(\mathbf u)$ is just  that functional evaluated over a discretization of space.  In this case, the nonlinear evolution of $\boldsymbol \psi(\mathbf u)$ is expressed as a superposition of fixed spatial structures (the Koopman modes) with an exponential time-dependence (through the Koopman eigenfunctions), giving the appearance of linearity. It is at present unclear how an expansion like (\ref{eqn:kmd}) might be constructed for a turbulent trajectory that visits multiple simple invariant solutions. For example, in a system with multiple equilibria, one would expect each fixed point to correspond to a neutral eigenfunction of the Koopman operator leading to a degeneracy of the $\lambda=0$ eigenvalue.

\subsection{Stuart-Landau equation}
For a simple example with multiple equilibria, we first revisit the problem considered in \citet{Bagheri2013} -- an analytical derivation of the Koopman decomposition for solutions to a Stuart-Landau equation, 
\begin{equation}
    \frac{\dd A}{\dd t} = a_0 A -a_1  A|A|^2.
\end{equation}
Following \citet{Bagheri2013}, we write the complex amplitude in polar coordinate, $A(t) = r(t)\text{exp}[\zi \theta(t)]$.
In our analysis we neglect the dependence on $\theta(t)$ and focus solely on the evolution of amplitude variable $r(t)$. 
The angular dependence is straightforward to incorporate and its inclusion only complicates the presentation. 
The evolution equation for $r(t)$ is
\begin{equation}
    \frac{\dd r}{\dd t} = \mu r - r^3
    \label{eqn:r_evoln}
\end{equation}
which has a pitchfork bifurcation at $\mu = 0$; for $\mu>0$ there are attractors at $r=\pm \sqrt{\mu}$ and a repellor at $r=0$. 
Similar to \citet{Bagheri2013}, we consider trajectories for which $r(t=0) >0$ and $r(t\to \infty) \to \sqrt{\mu}$ and seek a Koopman representation for an observable $\psi(r)$.
However, rather than inverse-engineering the Koopman eigenfunctions, eigenvalues and modes  from a Fourier expansion around the limit cycle (appendix A in \cite{Bagheri2013}), we identify them directly from the relationship (\ref{eqn:koop_ef}). Since this holds universally across the dynamics and not just close to any simple invariant solution, we can construct Koopman representations for the full lifespan of the solution trajectory. Interestingly, two different non-overlapping representations emerge, one centred around the repellor ($r=0$) and the other around the attractor ($r=\sqrt{\mu}$), which meet at a ``cross-over'' point where both fail simultaneously to converge.

\subsection{Koopman mode decompositions \label{KoopmanEx}}
Assuming $\mu > 0$, equation (\ref{eqn:r_evoln}) can be rescaled with $R:=\sqrt{\mu} r$ and $T:=\mu t$ to 
\begin{equation}
    \frac{\dd R}{\dd T} = R - R^3 =:f(R)
\label{scaled_1D_eqn}
\end{equation}
which has solution
\begin{equation}
    R(T; R_0) = \frac{1}{\sqrt{1 + b(R_0)e^{-2T}}},
    \label{eqn:exact_soln}
\end{equation}
where $b(R_0):= (1-R_0^2)/R_0^2$.
Our aim is to write the evolution of an observable, $\psi(R)$, as an expansion in eigenfunctions of the Koopman operator. For this one dimensional example, equation (\ref{eqn:koop_ef}) for the Koopman eigenfunctions becomes simply
\begin{equation}
    f(R) \frac{\dd \varphi_{\lambda}}{\dd R} = \lambda \varphi_{\lambda}.
\end{equation}
Hence 
\begin{equation}
    \varphi_{\lambda}(R) = \left(\frac{R^2}{1 - R^2}\right)^{\lambda/2} = \varphi_{\lambda}(R_0)e^{\lambda T}
    \label{eqn:efns}
\end{equation}
where $\lambda \in \mathbb R$ at least for analytic eigenfunctions (further restrictions will emerge below).
Equation (\ref{eqn:efns}) indicates that there is a single one-parameter family of Koopman eigenfunctions and a continuous spectrum of eigenvalues. A Koopman representation for a general observable would then be
\begin{align}
    \psi(R) &= \int_{-\infty}^{\infty}a_{\psi}(-\lambda) \varphi_{-\lambda}(R) \dd \lambda, \nonumber \\
    &= \int_{-\infty}^{\infty}a_{\psi}(-\lambda) \varphi_{-\lambda}(R_0)e^{-\lambda T} \dd \lambda
    \label{eqn:bilat}
\end{align}
where $a_{\psi}(-\lambda)$ is the \emph{Koopman mode density} for the observable $\psi$ corresponding to the Koopman eigenvalue $-\lambda$. Writing the integrand in terms of $-\lambda$ highlights the fact that the expression (\ref{eqn:bilat}) is a bilateral Laplace transform with $\lambda$ playing the role of the time-like variable and $T$ the transform variable. Setting $\psi=R$, which is often the first observable of interest, we write
\begin{equation}
    R(T) = \int_{-\infty}^{\infty}a(-\lambda) \varphi_{-\lambda}(R_0)e^{-\lambda T} \dd \lambda.
    \label{eqn:r_koop}
\end{equation}
and then the inverse Laplace transform inversion in the complex-$T$ plane
\begin{equation}
    a(-\lambda)\varphi_{-\lambda}(R_0) = \frac{1}{2\pi\zi}\int_{\gamma-\zi\infty}^{\gamma+\zi\infty}R(T)e^{\lambda T}\dd T
=\frac{1}{2\pi\zi}\int_{\gamma-\zi\infty}^{\gamma+\zi\infty}\frac{e^{\lambda T}}{\sqrt{1 + b(R_0)e^{-2T}}}\dd T
    \label{eqn:r_inv}    
\end{equation}
where $\gamma \in \mathbb R$ has to be chosen such that
\begin{equation}
\int^\infty_{-\infty} e^{-\gamma \lambda} |a(-\lambda)\varphi_{-\lambda}(R_0)| \dd \lambda < \infty.
\label{eqn:L1}
\end{equation}
For unilateral Laplace transforms, this just means choosing $\gamma$ to the right of all singularities in the complex transform variable plane. The convergence condition as  $\lambda \rightarrow -\infty$ is mute because $a(-\lambda)=0$ for all negative $\lambda$ (the time-like variable). For the bilateral Laplace transform,  the condition (\ref{eqn:L1}) becomes much more stringent. In particular, for $\lambda \rightarrow \infty$, $\gamma$ must be chosen to the right of all singularities in the $T$-plane (and the contour closed in the left hand plane) whereas for $ \lambda \rightarrow -\infty$, $\gamma$ must be to the left of all singularities in the $T$ plane (and the contour closed in the right hand plane). Clearly these are incompatible unless $a(\lambda)$ vanishes above or below some $\lambda_{crit}$. We now examine both possibilities.

%
%
The singularities of the integrand in (\ref{eqn:r_inv}) are the branch  points 
\begin{equation}
    T_n = \tfrac{1}{2}\text{ln}\,b + (n + \tfrac{1}{2})\zi \pi \qquad n \in {\mathbb Z}.
\end{equation}
and require branch cuts. Considering the case  of $a(\lambda)$ vanishing below some $\lambda_{crit}$,  these branch cuts are taken out to $-\infty$ parallel to the negative $\text{Re}(T)$ axis so that the Bromwich contour is closed to the left with a large semicircle, which is indented for each of the branch cuts. The contribution on the semicircle vanishes as its radius extends to infinity provided $\lambda > \lambda_{crit}:=-1$ so that the integral (\ref{eqn:r_koop}) reads
\begin{equation}
    R_+(T) = \int_{-1}^{\infty}a_+(\lambda)\varphi_{-\lambda}(R_0)e^{-\lambda T} \dd \lambda
    \label{eqn:r_att}
\end{equation}
which is a representation built upon Koopman eigenfunctions with eigenvalues in $(-\infty,1)$. The inverse Laplace transform (\ref{eqn:r_inv}) is a sum over keyhole contours, $C_n$, around the branch cuts 
\begin{equation}
    a_+(\lambda)\varphi_{-\lambda}(R_0) = -\frac{1}{2\pi \zi}\sum_{n=-\infty}^{\infty}\int_{C_n}\frac{e^{\lambda T}}{\sqrt{1 + be^{-2T}}}\dd T.
\end{equation}
Parameterising around each keyhole contour, it can be shown that
\begin{equation}
    a_+(\lambda)\varphi_{-\lambda}(R_0) = \frac{ (e ^{\zi \pi} b)^{  {\small {\lambda \over2} }}}{\pi} \int_0^{\infty}(u^2 + 1)^{-{\small{\lambda \over2}} - 1} \dd u \sum_{n=-\infty}^{\infty}e^{\zi \pi \lambda n}.
\end{equation}
The infinite sum of complex exponents is the Fourier representation of a Dirac comb, $\sum_n e^{\zi \pi \lambda n} = 2\sum_n \delta(\lambda - 2n)$.
After dividing by the eigenfunction $\varphi_{-\lambda}$, the Koopman mode density is found to be
\begin{equation}
    a_+(\lambda) = \frac{2e^{\zi \pi {\small {\lambda \over2}}}}{\pi} \underbrace{\int_0^{\infty}(u^2 + 1)^{-{\small {\lambda\over2}} - 1} \dd u}_{I(\lambda)} \sum_{n=-\infty}^{\infty}\delta(\lambda -2n).
\end{equation}
Since $\lambda > -1$ and $\lambda$ is even, only Koopman eigenfunctions not associated with exponential growth are included in the representation. 
Moreover, for discrete $\lambda_n \in \{0, 2, 4, \dots\}$, it can be shown that $I(\lambda_n) = (1-\tfrac{1}{2n})I(\lambda_{n-1})$, with $I(\lambda_0)=\tfrac{\pi}{2}$, so
\begin{equation}
    I(\lambda_n) = \frac{(2n)!}{2^{2n}(n!)^2}\frac{\pi}{2}.
\end{equation}
Use of the Koopman mode density $a_+(\lambda)$ in (\ref{eqn:r_att}) thus picks out a discrete Koopman expansion around the attractor $R=1$ as found earlier by \citet{Bagheri2013},
\begin{equation}
    R_+(T) = \sum_{n=0}^{\infty}  \underbrace{\frac{(-1)^n(2n)!}{2^{2n}(n!)^2}}_{\hat{R}_{-2n}}\varphi_{-2n}(R_0) e^{-2n T}
    \label{eqn:exp_att}
\end{equation}
where $\hat{R}_{-2n}$ is the Koopman mode for the observable $\psi=R$ associated with the Koopman eigenfunction $\varphi_{-2n}$. This series can be recognised as just the Taylor expansion of the exact solution (\ref{eqn:exact_soln})
\begin{equation}
R(T; R_0) = \frac{1}{\sqrt{1 + \tfrac{1-R_0^2}{R_0^2}e^{-2T}}}
\label{eqn: exact_y}
\end{equation}
in $y:=\tfrac{1-R_0^2}{R_0^2}$ about $y=0$ valid for dynamics `close' to the attracting fixed point $R=1$ (\cite{Bagheri2013} proceeded in the opposite direction starting from the Taylor expansion to  deduce the Koopman expansion).  Since $\tfrac{1-R(T)^2}{R(T)^2}=\tfrac{1-R_0^2}{R_0^2}e^{-2T}$, we can rewrite this as the identity
\begin{equation}
R=\frac{1}{\sqrt{1 + \tfrac{1-R^2}{R^2}}}
\label{eqn:id}
\end{equation}
which indicates that the representation (\ref{eqn:exp_att}) will fail to converge for $(1-R^2)/R^2 \geq 1$ or $R\leq 1/\sqrt{2}$. In other words, since $R(T)$ increases monotonically with time, the representation (\ref{eqn:exp_att}) holds for any solution with initial condition $R_0> 1/\sqrt{2}$.

\begin{figure}
    \centering
    \includegraphics[width=0.47\textwidth]{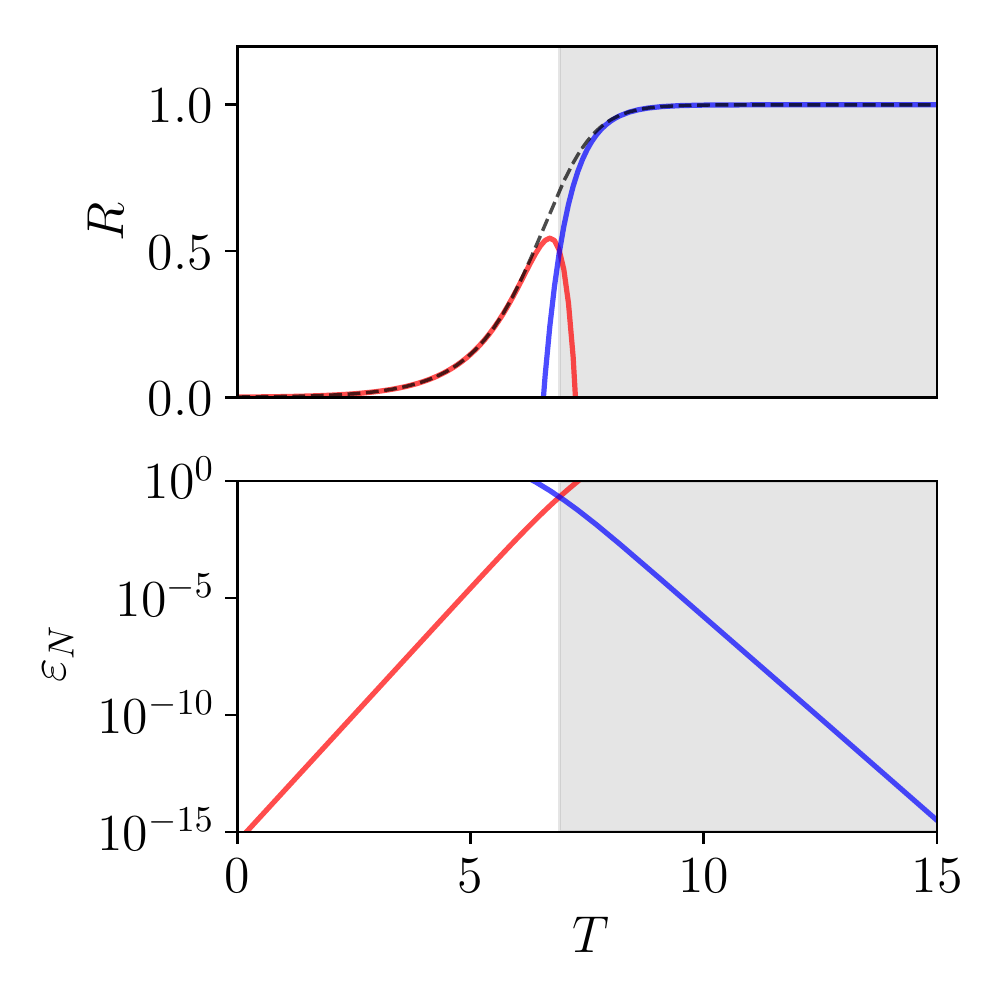}
    \includegraphics[width=0.47\textwidth]{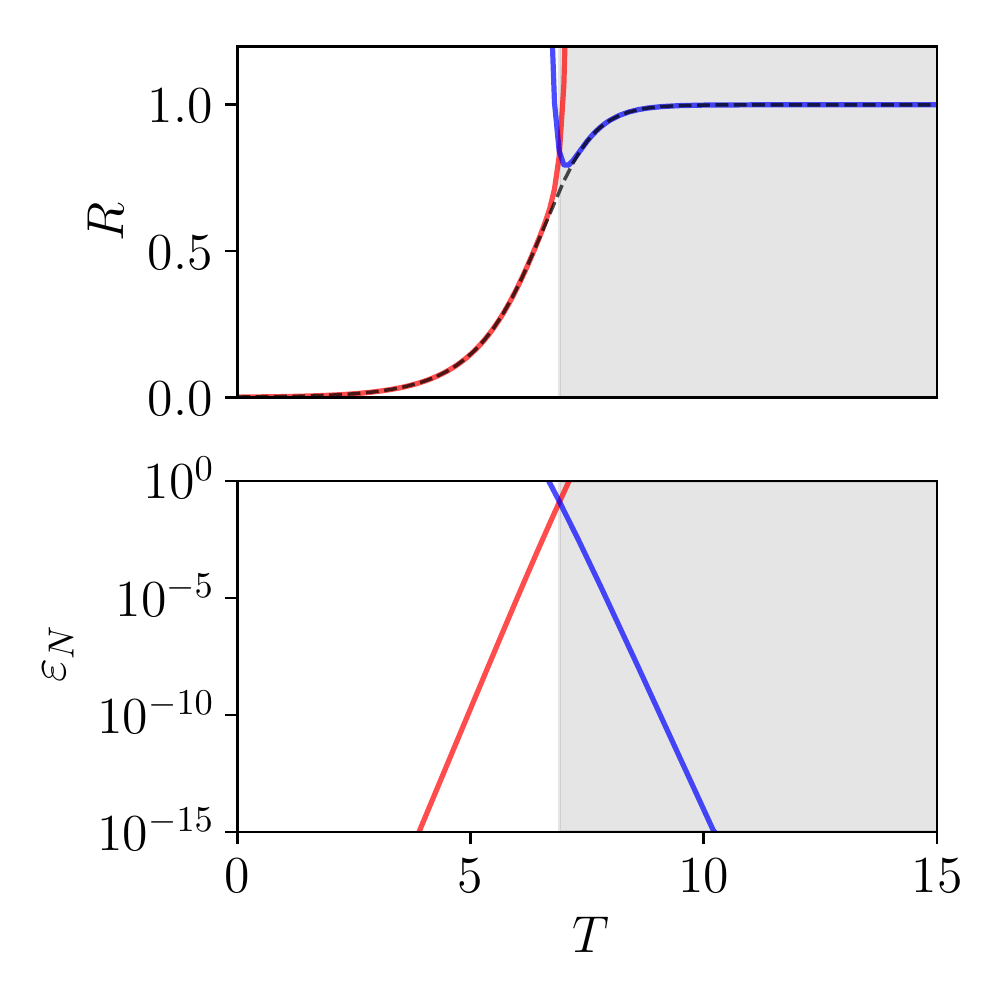}
    \caption{(Top) True evolution (\ref{eqn:exact_soln}) with $R_0=10^{-3}$ (dashed black) with Koopman approximations (\ref{eqn:exp_att} and \ref{eqn:exp_rep}) overlaid (left: $N=2$ modes, right: $N=5$; red/blue correspond to the attracting/repelling expansions respectively). (Bottom) The error $\varepsilon_N:=|R(t)-R_{\pm}(t;N)|$, where $N$ is the number of modes included in the expansion. Grey region identifies the crossover point betweeen repelling and attracting Koopman expansions.}
    \label{fig:err}
\end{figure}

%
%
We now turn our attention to the other possible scenario where  $a(\lambda)$ vanishes {\em above} some $\lambda_{crit}$. The branch cuts must now be taken out to $+\infty$ parallel to the positive $\text{Re}(T)$ axis so that the Bromwich contour is closed to the {\em right} with a large semicircle, which is indented for each of the branch cuts. The contribution on the semicircle vanishes as its radius extends to infinity provided that $\lambda < \lambda_{crit}:=0$ and we write
\begin{equation}
    R_-(T) = \int_{-\infty}^{0}a_-(\lambda)\varphi_{-\lambda}(R_0)e^{-\lambda T} \dd T
    \label{eqn:r_rep}
\end{equation}
so only Koopman eigenfunctions associated with exponential growth are included. Parameterising around the branch cuts as previously yields the  Koopman mode density,
\begin{equation}
    a_-(\lambda) = \frac{2e^{-  {\small {1\over2}} \zi\pi (\lambda+1) }}{\pi} \int_0^{\infty}(u^2+1)^{ {\small {\lambda-1 \over2} }}\dd u \sum_{n=-\infty}^{\infty}\delta(\lambda+1-2n).
\end{equation}
As before, we can  evaluate  the integral using a recurrence relation and using (\ref{eqn:r_rep}) we recover another Koopman expansion 
\begin{equation}
   R_-(T) = \sum_{n=0}^{\infty}  \frac{(-1)^n(2n)!}{2^{2n}(n!)^2}\varphi_{2n+1}(R_0) e^{(2n+1) T}.
    \label{eqn:exp_rep}
\end{equation}
This is the Taylor expansion in $z:=\tfrac{R}{\sqrt{1-R^2}}$ around $z=0$ of the exact solution
\begin{equation}
R=\frac{z}{\sqrt{1+z^2}}
\end{equation}
(a simple manipulation of the identity (\ref{eqn:id})) which fails to converge when $z=1$ or $R(T) \geq 1/\sqrt{2}$. So if $R_0 < 1/\sqrt{2}$, this representation will hold until $R=1/\sqrt{2}$. Beyond this point in time, the other Koopman expansion can then be used to represent the solution. So the two Koopman decompositions (\ref{eqn:exp_att} and \ref{eqn:exp_rep}) together allow (almost) the entire nonlinear evolution to be expressed as a superposition of linear (exponential time dependence) observables. The performance of the two decompositions, truncated at a finite number of Koopman modes, is examined in figure \ref{fig:err}. As expected, the two expansions fail as they are pushed beyond the crossover point $R=1/\sqrt{2}$.

At this point it is interesting to ask what goes wrong in attempting to build a Koopman expansion centred around another point (say, even $R=1/\sqrt{2}$) which is not an equilbrium. Here a connection with Carleman linearization \citep{Carleman1932} is useful.  Carleman linearization makes a nonlinear system linear by relabelling each nonlinearity as a new dependent variable of the system. Typically, this converts a finite dimensional nonlinear system into an infinite linear system as additional equations need to be added to describe how the new dependent variables evolve. This  generically introduces further nonlinearities and the procedure mushrooms with yet more variables needing to be defined. When this linearization procedure is carried out around a solution of the system such as an equilibrium, it produces a purely linear system as opposed to the generic affine one - i.e. the time evolution of the system is given by  a linear operator. The (adjoint) eigenfunctions and eigenvalues (modulo exponentiation)  of this operator would then seem to  correspond with the Koopman (eigenfunctions) modes and eigenvalues of the Koopman operator. A Koopman expansion centred at this point clearly makes sense. In contrast, if the Carleman linearization is performed around a non-solution, the resulting system is then only affine and the temporal evolution cannot be purely expressible as a sum of exponentially time varying Koopman modes: see the Appendix for details for the model studied here. Thus, it would only seem to make sense to talk about Koopman expansions about simple invariant solutions or just  the equilibria $R=\{0,1\}$ here for $R \geq 0$.

Finally, it is worth emphasizing that the breakdown of a given Koopman expansion is associated with a loss of convergence rather than any  pathology in the component Koopman eigenfunctions.  In fact, the Koopman eigenfunctions exist everywhere away from  the fixed points $R=\{0,1\}$. The point is just that certain subsets can't be used in a convergent representation at a given point in the dynamics. 
The jump between Koopman expansions has important consequences for DMD, which we now explore.

%
%
\subsection{Dynamic mode decomposition}
In his examination of the transient collapse of the flow over a cylinder onto the vortex-shedding limit cycle, \citet{Bagheri2013} noted a curious dependence of the output of DMD on the length of time window over which data was collected.
If the observation window was restricted to a time interval where the velocity, $\mathbf u$, is `close' to the periodic orbit, DMD accurately reproduced the Koopman eigenvalues for the attracting expansion.
However, when the window was extended to also include the `amplification' region, the DMD eigenvalues appeared as discrete approximations to continuous lines of decaying eigenvalues.
This phenomenon is a consequence of the jump between Koopman expansions around the fixed points of equation (\ref{eqn:r_evoln}).

In fact, the crossover point between Koopman expansions has a critical impact on the ability of DMD to approximate a Koopman expansion even if the DMD design is perfect in all other respects -- i.e. the elements of the user-defined observable vector, $\boldsymbol \psi(\mathbf u)$, are a suitable basis for the Koopman eigenfunctions and a sufficiently large amount of data is examined \citep{Williams2015}.
In particular, if the desire is to obtain a Koopman decomposition of the state variable, the DMD must be restricted to within a neighbourhood of an exact solution inside which the expansion is valid.
We demonstrate this behaviour by conducting DMD of the simple 1D problem (\ref{eqn:r_evoln}), successfully obtaining Koopman eigenvalues only when snapshot pairs are restricted to times $\{T: R(T)<1/\sqrt{2}\}$ or $\{T: R(T)>1/\sqrt{2}\}$, but not for an overlapping interval. 

%
%
\begin{figure}
    \centering
    \includegraphics[width=0.3\textwidth]{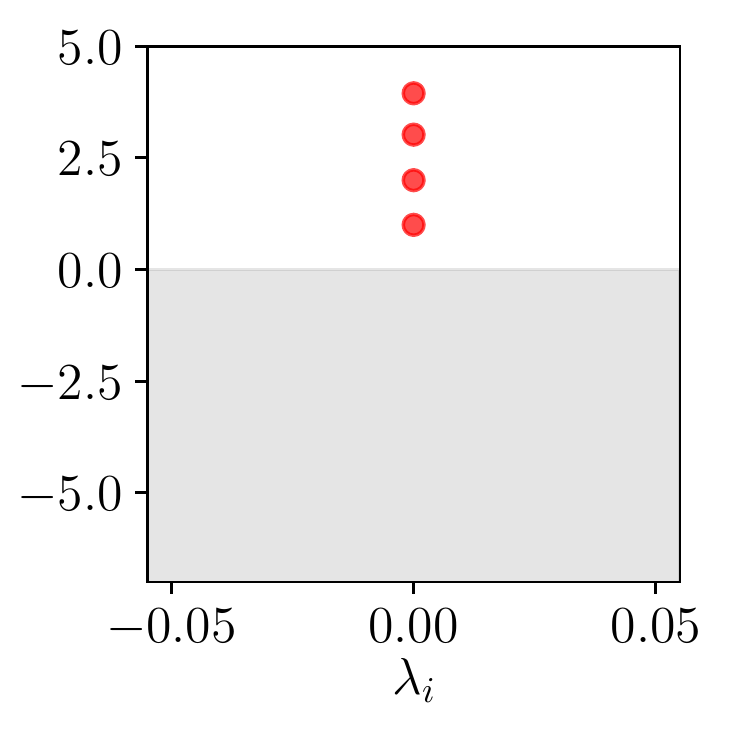}
    \includegraphics[width=0.3\textwidth]{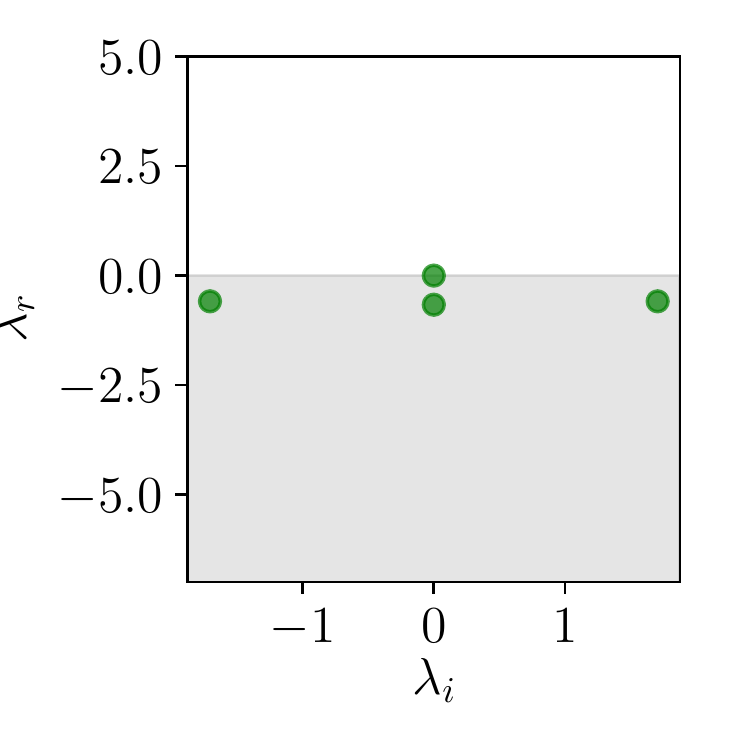}
    \includegraphics[width=0.3\textwidth]{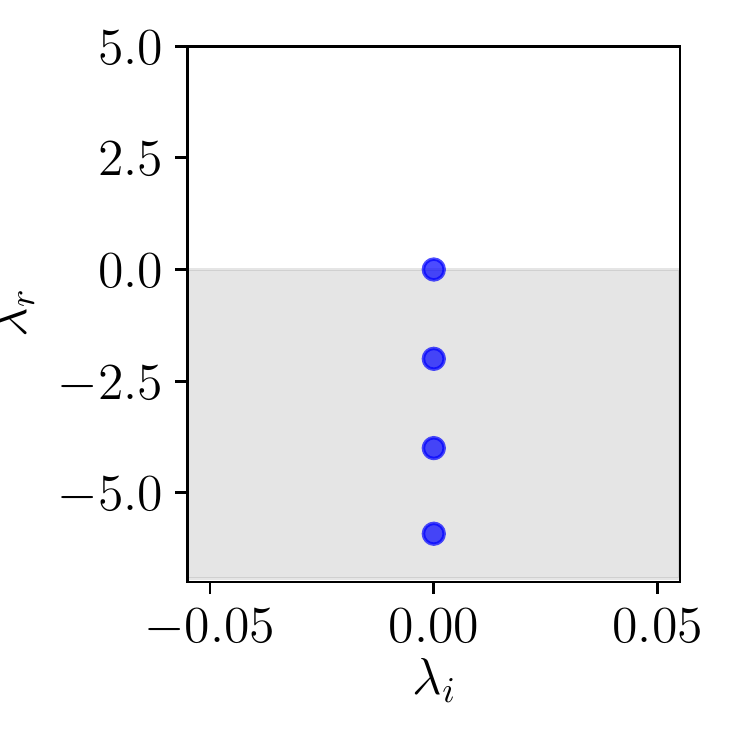}
    \caption{Eigenvalues obtained from DMD on the evolution shown in figure \ref{fig:err} with $\boldsymbol \psi(R) = (R,R^2,R^3,R^4)^T$. $M=20$ snapshots and $\delta t=1$ obtained on (left) $t\in [0,5)$, (centre) $t\in [5,15)$ and (right) $t\in[10, 20)$.}
    \label{fig:dmd_eigs}
\end{figure}
We generate snapshots $\{R_i\}$ of the trajectory reported in figure \ref{fig:err} with a spacing $\Delta t_s = 0.1$ on the interval $t \in [0,20]$.
For the DMD, we use a snapshot spacing $\delta t = 1$, so in total we have available $M_{\text{max}}=190$ snapshot pairs (\,t=0 to 19 step 0.1 mapped to t=1 to 20 step 0.1\,). 
The observable vector for the DMD is made up of polynomials in $R$,
$\boldsymbol \psi (R) = (R, R^2, R^3, R^4)$.
The DMD reported here is slightly unusual in the sense that $N$, the dimension of $\boldsymbol \psi$, is much less than $M$, the number of snapshots.
Typically $N \gg M$ in fluid mechanics, and it will be shown in \S3 that analogous behaviour to that found in this 1D problem occurs along heteroclinic connections between equilibria of the Navier-Stokes equations.

The DMD methodology is essentially as specified in \citet{Tu2014}, although the inclusion of polynomials of the state in $\boldsymbol \psi$ makes the current problem an example of EDMD \citep{Williams2015}.
Given a matrix of snapshots,
\begin{equation}
    \boldsymbol \Psi^t = \begin{bmatrix} \boldsymbol \psi(R(t_i)) & \boldsymbol \psi(R(t_j)) & \cdots & \end{bmatrix},
\end{equation}
and a corresponding matrix with the observables now evaluated $\delta t$ later,
\begin{equation}
    \boldsymbol \Psi^{t+\delta t} = \begin{bmatrix} \boldsymbol \psi(R(t_i+\delta t)) & \boldsymbol \psi(R(t_j+\delta t)) & \cdots & \end{bmatrix},
\end{equation}
the DMD operator $\hat{\mathbf K}$ is the linear operator which best maps between corresponding snapshot pairs (in a least squares sense), 
\begin{equation}
    \hat{\mathbf K} := \boldsymbol \Psi^{t+\delta t}(\boldsymbol \Psi^t)^+,
\end{equation}
where the $+$ superscript indicates a pseudo (Moore-Penrose) inverse.
Note that the snapshot times, $\{t_i\}$, do not need to be sequential, and are drawn randomly from within the time interval of interest.
As described in \citet{Rowley2017}, the right eigenvectors of the DMD operator, $\hat{\boldsymbol \psi}_j=(\hat{R}_j, \hat{R}^2_j, \hat{R}^3_j, \hat{R}^4_j)^T$, approximate Koopman modes, while the left eigenvectors, $\mathbf w_j$, can be used to find the Koopman eigenfunctions,
\begin{equation}
    \varphi_j(R) = \mathbf w_j^H \boldsymbol \psi(R),
\end{equation}
under the assumptions that (i) the elements of $\boldsymbol \psi$ constitute a suitable basis for the eigenfunctions and (ii) sufficient data has been collected such that $\mathbf w \in \text{range}(\boldsymbol \Psi^t)$.

Eigenvalues from three DMDs are reported in figure \ref{fig:dmd_eigs}.
Each calculation was performed on snapshot pairs extracted from a different time window.
When the time window is limited to the repelling region, DMD yields eigenvalues $\lambda_n = n$ (while the expansion for $R$ around the repellor requires only odd integers, the inclusion of powers of $R$ in the observable means a larger set, $n \in \mathbb N$, are uncovered: odd integers can sum to be  even). 
When the time window lies within the region of validity for the attracting expansion, DMD finds the attractor eigenvalues, $\lambda_n=-2n$ (sums of even integers remain even). 
On the other hand, DMD on snapshots from a time window which overlaps both expansion regions is unable to find eigenvalues for either expansion. 

%
%
\begin{figure}
    \centering
    \includegraphics[width=0.24\textwidth]{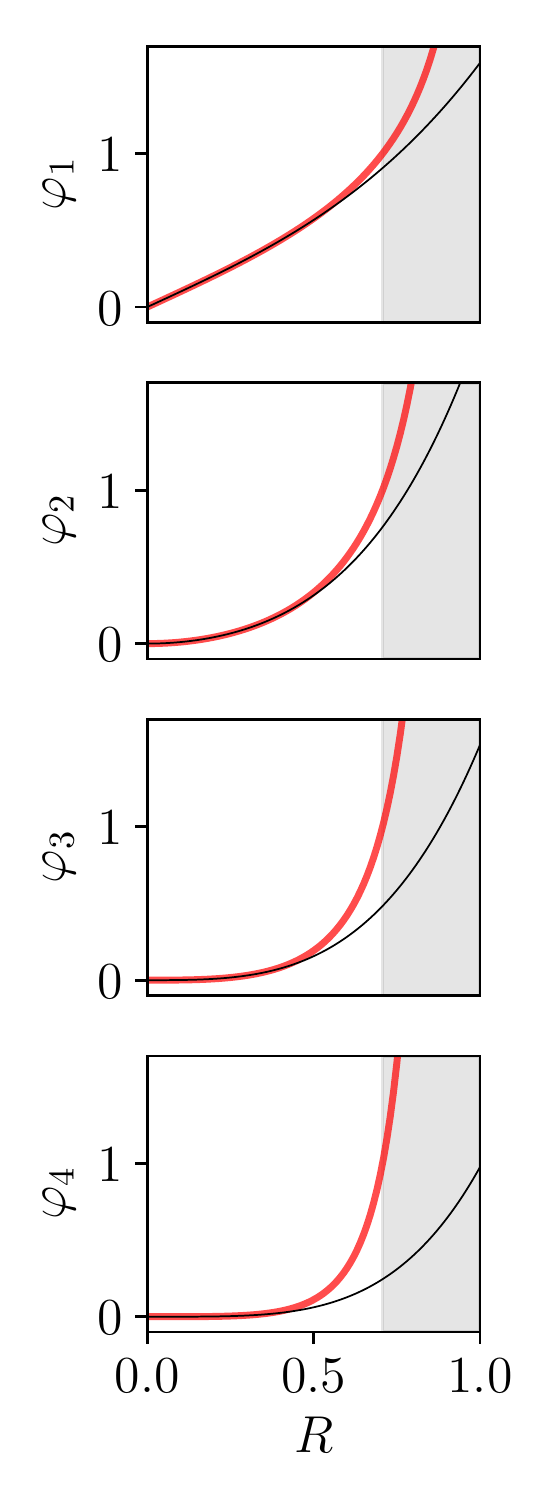}
    \includegraphics[width=0.24\textwidth]{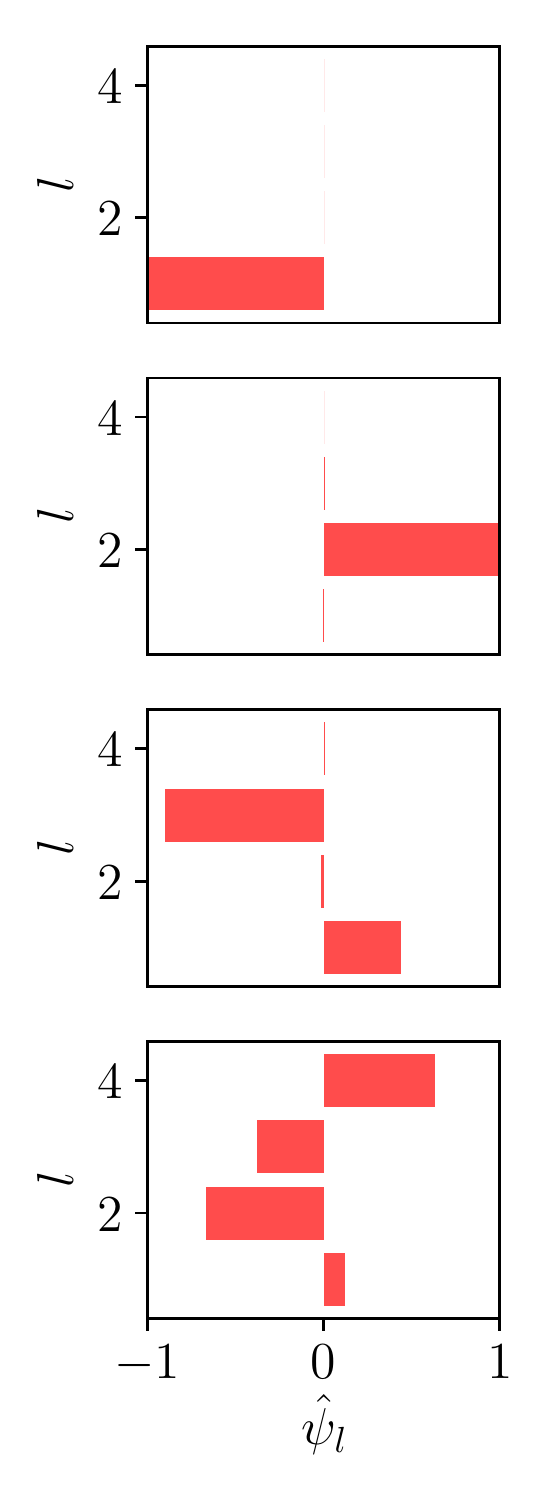}
    \includegraphics[width=0.24\textwidth]{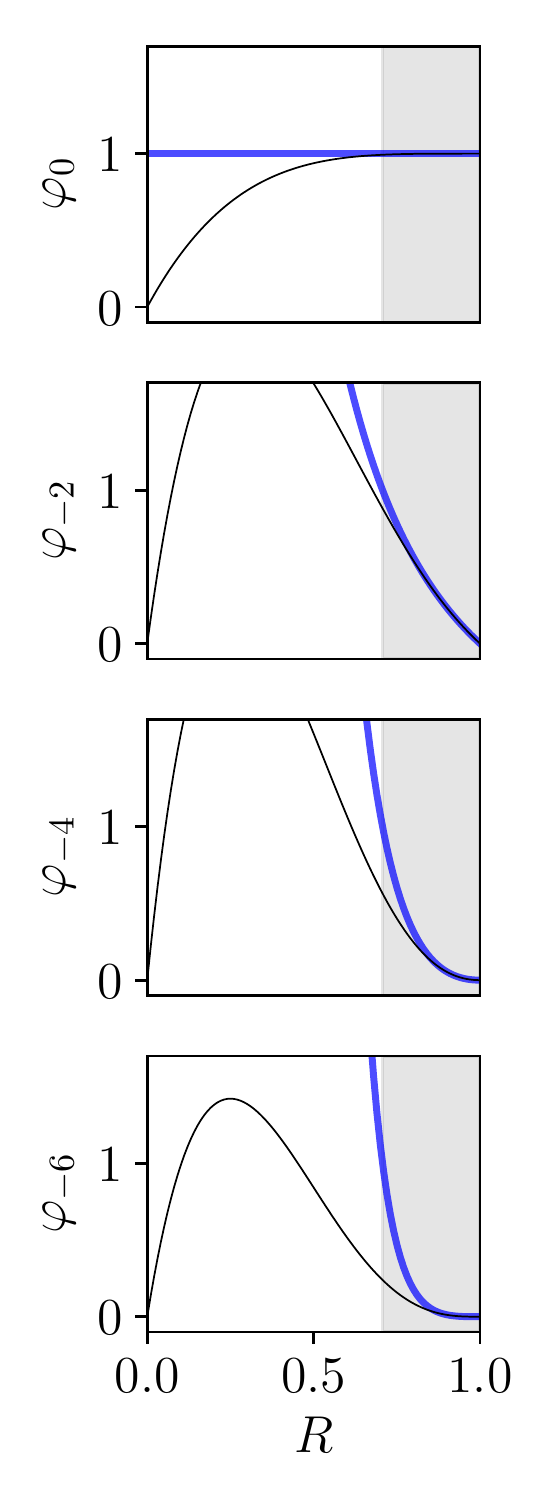}
    \includegraphics[width=0.24\textwidth]{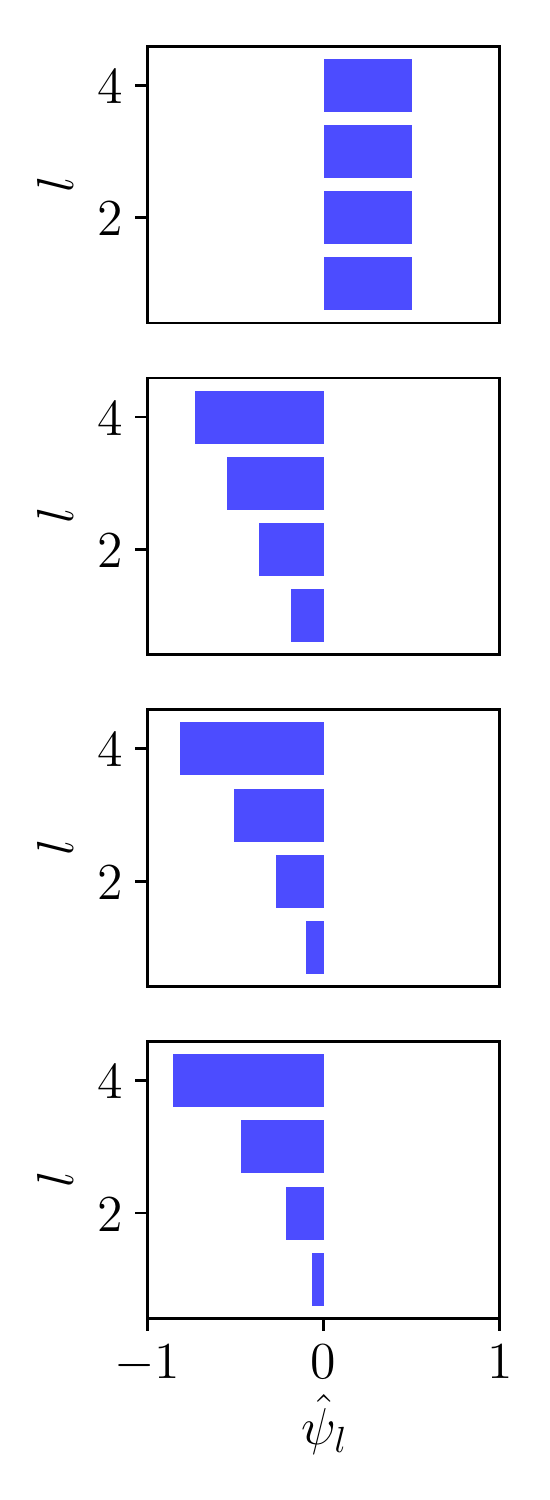}
    \caption{Comparison of Koopman eigenfunctions (colours) with eigenfunctions extracted from the repelling and attracting DMDs (black) reported in figure \ref{fig:dmd_eigs}, alongside corresponding DMD modes $\hat{\boldsymbol \psi}$. Note that the $l^{th}$ component of the the $j^{th}$ DMD mode represents the DMD approximation to $\hat{R}^l_j$, the $j^{th}$ Koopman mode in the Koopman decomposition of $R^l$.}
    \label{fig:dmd_eigfs}
\end{figure}
The performance of the DMD can be assessed in more detail by comparing the predicted Koopman eigenfunctions and modes to those derived in \S\ref{sec:SL}.
In figure \ref{fig:dmd_eigfs} the DMD approximations to the Koopman eigenfunctions are reported for both the ``repelling'' and ``attracting'' windows.
In both cases, the DMD algorithm is able to build a locally valid approximation to the true Koopman eigenfunction from the polynomials $R^m$ in the observable vector $\boldsymbol \psi$.
These locally valid expansions break down as the crossover point, $R=1/\sqrt{2}$, is approached.
The correspondence between DMD and Koopman also gets progressively worse for the higher order eigenfunctions -- a consequence of the limited number of polynomials in $\boldsymbol \psi$.

The DMD modes reported in figure \ref{fig:dmd_eigfs} should be interpreted in the following way:
The $l^{th}$ component of the DMD mode alongside eigenfunction $\varphi_j$ is the DMD approximation to the $j^{th}$ Koopman mode in an expansion of $R^l$ i.e. $\hat{R}^l_j$.
So, for example, component $\hat{\psi}_{l=2}$ alongside eigenfunction $\varphi_4$ (bottom left corner of figure \ref{fig:dmd_eigfs}) is the DMD approximation to Koopman mode $\hat{R}^2_4$ in the expansion $R^2 = \sum_{m\in \mathbb N} \varphi_{2m}(R)\hat{R}^2_{2m}$.
The DMD approximations to the Koopman modes reported in figure \ref{fig:dmd_eigfs} are consistent with the analytical expansions derived in \S\ref{sec:SL}.
For example, the repellor decomposition (\ref{eqn:exp_rep}) indicates that the Koopman eigenvalues required to advance $R(T)$ are the odd integers.
The DMD identifies a broader set of Koopman eigenvalues, $\lambda \in \mathbb N$, than those needed for $R$ alone, but correctly finds that the Koopman mode $\hat{R}_2=0$ while picking up the contributions $\hat{R}_1$ and $\hat{R}_3$ (the DMD mode for $\hat{R}_4$ is non-zero but small -- DMD with higher order polynomials included in $\boldsymbol \psi$ can eliminate this error). 

The first non-zero Koopman eigenvalue in both expansions is the growth/decay rate associated with the locally linear dynamics around the repelling and attracting equilibria respectively.
The higher order terms in the Koopman decompositions allow us to propagate observables (in particular the state variable itself) beyond these linear subspaces, and we have demonstrated here that DMD is a robust method for finding these contributions provided that the observation window is contained within a particular ``expansion region''. 
In the remainder of this paper we will show how similar behaviour is observed along heteroclinic connections between equilibria of the Navier-Stokes equations, and that DMD can successfully identify modes associated with repelling and attracting expansions along their unstable and stable manifolds, respectively.

%
%
%
%
\section{Heteroclinic connections in plane Couette flow}
\label{sec:HC}
In this section we use DMD to search for crossover points between simple invariant solutions of the Navier-Stokes equations.  
The flow configuration is Couette flow with no-slip boundary conditions at the top and bottom walls and periodic boundary conditions in both horizontal directions. 
The problem is non-dimensionalised by the channel half-height, $d$, and the plate velocity $U_0$ (so the boundary conditions become ${\mathbf u}(x,y,\pm 1,t)=\pm {\bf \hat{x}}$), leading to a Reynolds number $Re:= U_0 d/\nu$.  

The Navier-Stokes equations are solved using a fractional-step method in which the diffusion terms are treated implicitly with Crank-Nicholson and an explicit third-order Runge-Kutta scheme is used for the advection terms.  Spatial discretisation is performed with second-order finite differences on a staggered grid. 
The code is wrapped inside a Newton-GMRES-Hookstep algorithm \citep[e.g.][]{Viswanath2007,Gibson2008,Chandler2013} that can be used to converge equilibria and (relative) periodic orbits, and has been validated by reproducing many known equilibria and periodic orbits in both the `GHC' box of \citet{Gibson2008} and the `HKW' box of \citet{Hamilton1995}.

\subsection{Heteroclinic connection between Nagata solutions}
%
%
\begin{figure}
    \includegraphics[width=0.49\textwidth]{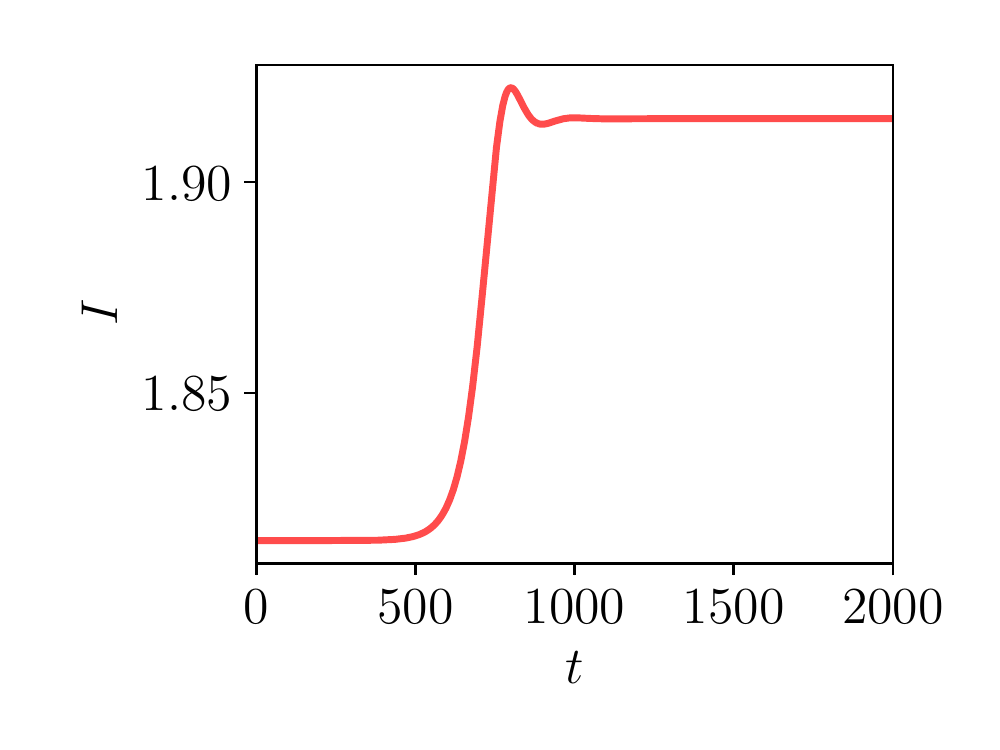}
    \includegraphics[width=0.47\textwidth]{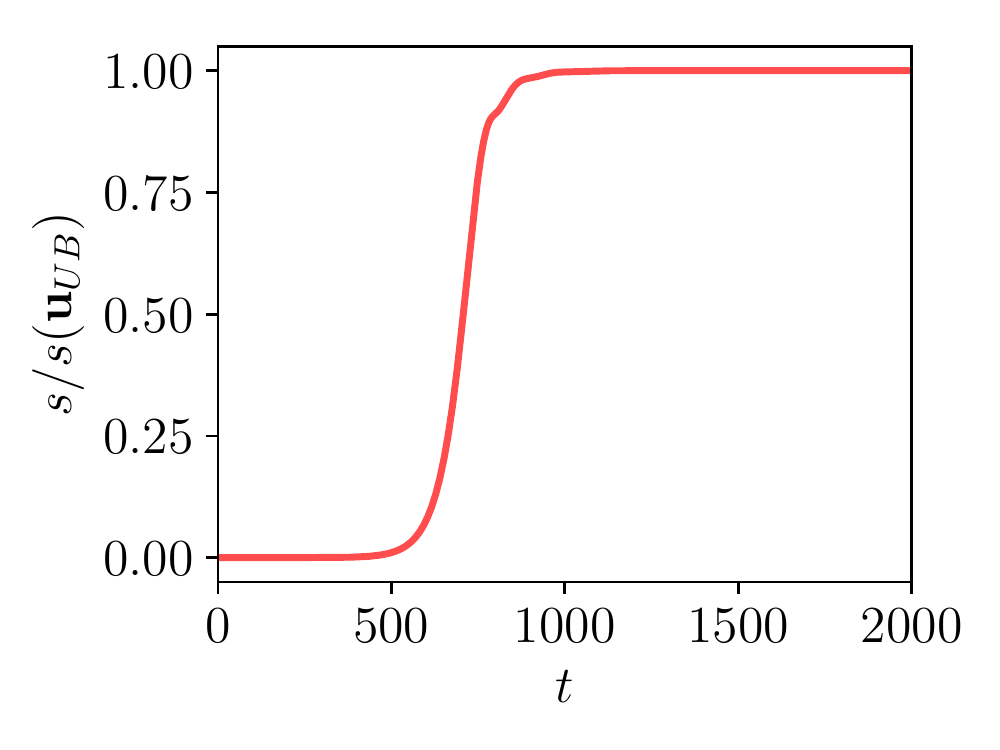}
    \caption{(Left) Energy production along finite-time approximation to the $\mathbf u_{LB}\to \mathbf u_{UB}$ heteroclinic connection at $Re=135$. (Right) Arclength along the heteroclinic connection measured as distance from $\mathbf u_{LB}$.}    \label{fig:Re135_It}
\end{figure}

We consider a \citet{Nagata1990} box of size $(L_x, L_y, L_z)=(5\pi/2, 4\pi/3, 2)$ initially at $Re=135$.
In this configuration the Navier-Stokes equations support three equilibrium solutions -- the constant shear solution $\mathbf u_C$ and the Nagata lower- and upper-branch solutions,  $\mathbf u_{LB}$ and $\mathbf u_{UB}$ respectively \citep{Nagata1990}. 
These two solutions are born out of a saddle-node bifurcation at around $Re\sim 125$ \citep{Nagata1990} for this box. 
At $Re=135$ both $\mathbf u_C$ and $\mathbf u_{UB}$ are stable while $\mathbf u_{LB}$ is the unstable edge state on the dividing manifold between their respective basins of attraction.  We compute $\mathbf u_{LB}$ in the `GHC' box at $Re=400$ by using a snapshot of (transient) turbulence as a guess in the Newton-GMRES-hookstep algorithm described above. This solution is then continued down to the target Reynolds number and target box size.
A finite-time approximation to the heteroclinic connection between $\mathbf u_{LB}$ and $\mathbf u_{UB}$ is then obtained in the following manner:
(i) velocity snapshots are generated along a short trajectory $t \in [0,50]$ with the initial condition $\mathbf u_0 = (1+\varepsilon)\mathbf u_{LB} - \varepsilon \mathbf u_C$, where $\varepsilon = 10^{-6}$;
(ii) the unstable eigenfunction, $\hat{\mathbf u}_1$, is extracted from this trajectory using DMD;
(iii) the new initial condition $\mathbf u_0' = \mathbf u_{LB} + \delta \hat{\mathbf u}_1$, where $\delta|\mathbf u_{LB}| = 10^{-8}$ is then used to compute a more accurate approximation to the heteroclinic connection.   
At $Re=135$, the first initial condition $\mathbf u_0$ is actually sufficient to obtain a good approximation to the heteroclinic connection since the upper branch solution is stable. However, at higher Reynolds numbers $\mathbf u_{UB}$ becomes unstable, and the initial condition described in (iii) can generate trajectories which still spend some time in its vicinity before being flung out along its unstable manifold.

The energy production, 
\begin{equation}
I':=\frac{1}{2L_xL_y}\int^{L_x}_0 \!\!\int^{L_y}_0  
\frac{1}{Re}
\frac{\partial u}{\partial z}\biggl|_{z=\pm 1}\biggr.
\! \!\dd x \dd y
\end{equation}
per unit area and arclength and normalized by its value in laminar flow ($I:=I'/I_{lam}'$) is computed along the heteroclinic connection and is reported as a function of time in figure \ref{fig:Re135_It} to highlight
the qualitative similarity with the evolution $R(T)$ in the model problem of \S2. 
Similar to the behaviour near to the origin there, the Nagata lower branch solution is a repellor with a single unstable direction. However, the stable subspace around the attractor, $\mathbf u_{UB}$, is more complex (four dimensional), as described below.
While analytical construction of Koopman decompositions around these fixed points is not possible here, we employ DMD to identify repelling and attracting Koopman expansions.

%
%
\begin{figure}
    \centering
    \includegraphics[width=0.8\textwidth]{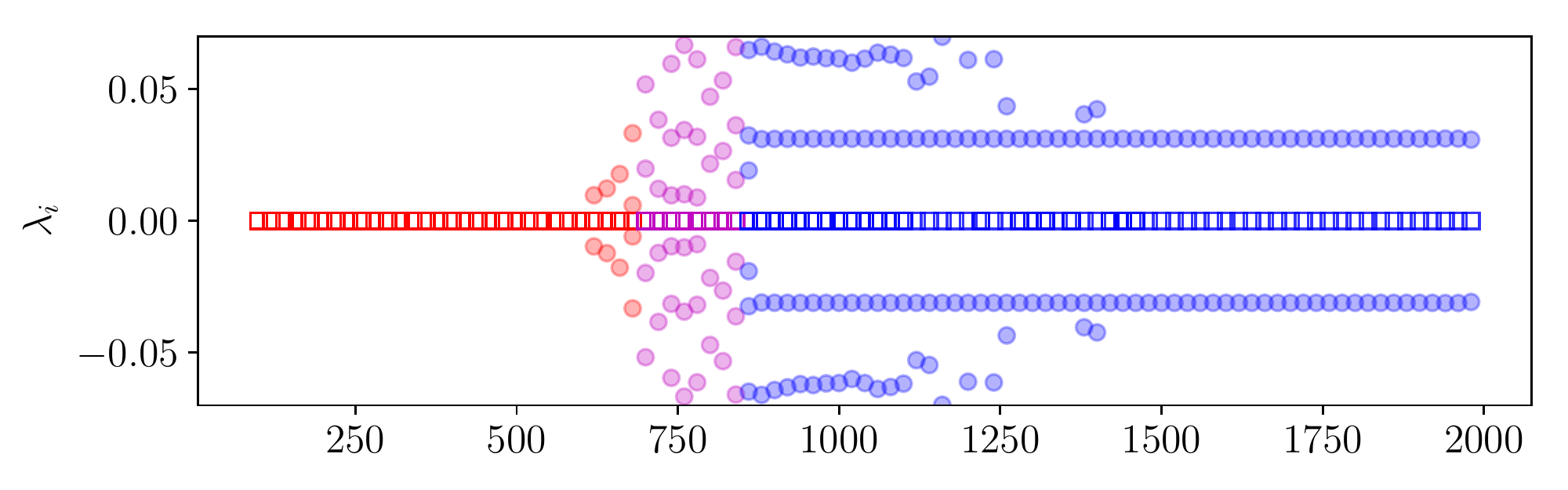}
    \includegraphics[width=0.8\textwidth]{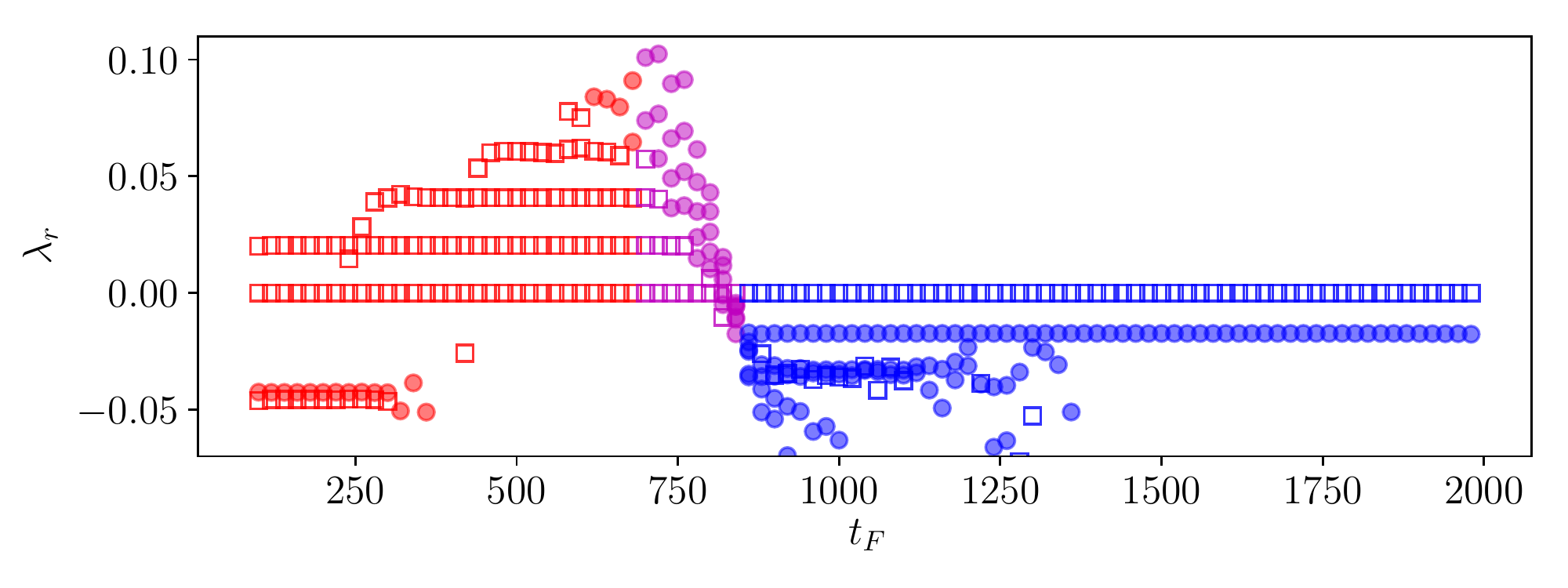}
    \caption{(Top) Real and (bottom) imaginary components of DMD eigenvalues obtained for a time window of length $T_w=100$ passed through the trajectory shown in figure \ref{fig:Re135_It} (the variable $t_F$ is the ``final'' time of each time window). Each individual calculation is performed on $M=25$ snapshots pairs separated by $\delta t=2$ selected randomly within each time window. Red and blue colouring indicate whether the behaviour is classified as locally ``repelling'' or ``attracting'' respectively.}
    \label{fig:dmd_window}
\end{figure}
For the DMD, 1000 snapshots of the full velocity field, separated by $\Delta t_s=2$, are stored for the trajectory in figure \ref{fig:Re135_It} and the observable vector is 
\begin{equation}
    \boldsymbol \psi(\mathbf u) = \mathbf u - \mathbf u_C.
\end{equation}
Initially, we pass a fixed time window of width $T_w=100$ along the heteroclinic connection, performing many DMD calculations with the results collated in figure \ref{fig:dmd_window}.
Each individual calculation is performed with $M=25$ snapshot pairs separated by $\delta t=2$ extracted randomly from within the interval of interest. 
Initially, and as anticipated for a trajectory repelled from the edge, the DMD identifies a single unstable eigenvalue $\lambda_1 \approx 0.02$ associated with the unstable linear subspace about $\mathbf u_{LB}$.
As the time window is passed along the heteroclinic connection, further unstable eigenvalues $\lambda_n = n\lambda_1$ are uncovered.
This suggests that the DMD algorithm is identifying Koopman eigenfunctions in the same family as $\varphi_{\lambda_1}(\mathbf u)$, i.e. $\varphi_{\lambda_n}(\mathbf u) = \varphi_{\lambda_1}^n(\mathbf u)$ (higher harmonics of the primary instability) in analogy to the model problem considered in \S2.

%
%
\begin{figure}
    \centering
    \includegraphics[width=0.55\textwidth]{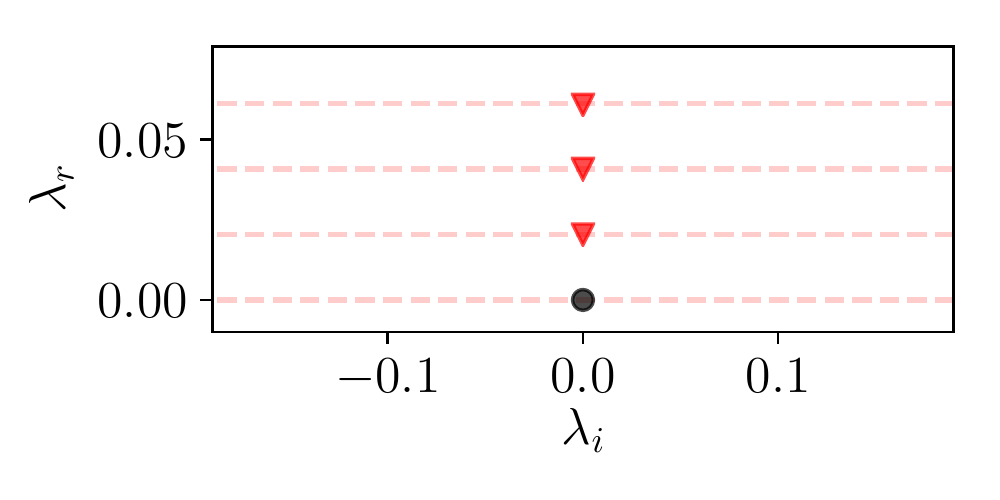}
    \caption{Eigenvalues obtained from a DMD calculation on $t\in[400, 500]$ from the trajectory shown in figure \ref{fig:Re135_It} (cf figure \ref{fig:dmd_window}). 
    The number of snapshots is $M=25$ and $\delta t=2$. The dashed red lines identify integer multiples of the first unstable mode, $\lambda_1\approx 0.020$.}
    \label{fig:growth_eigs}
    \includegraphics[width=0.52\textwidth]{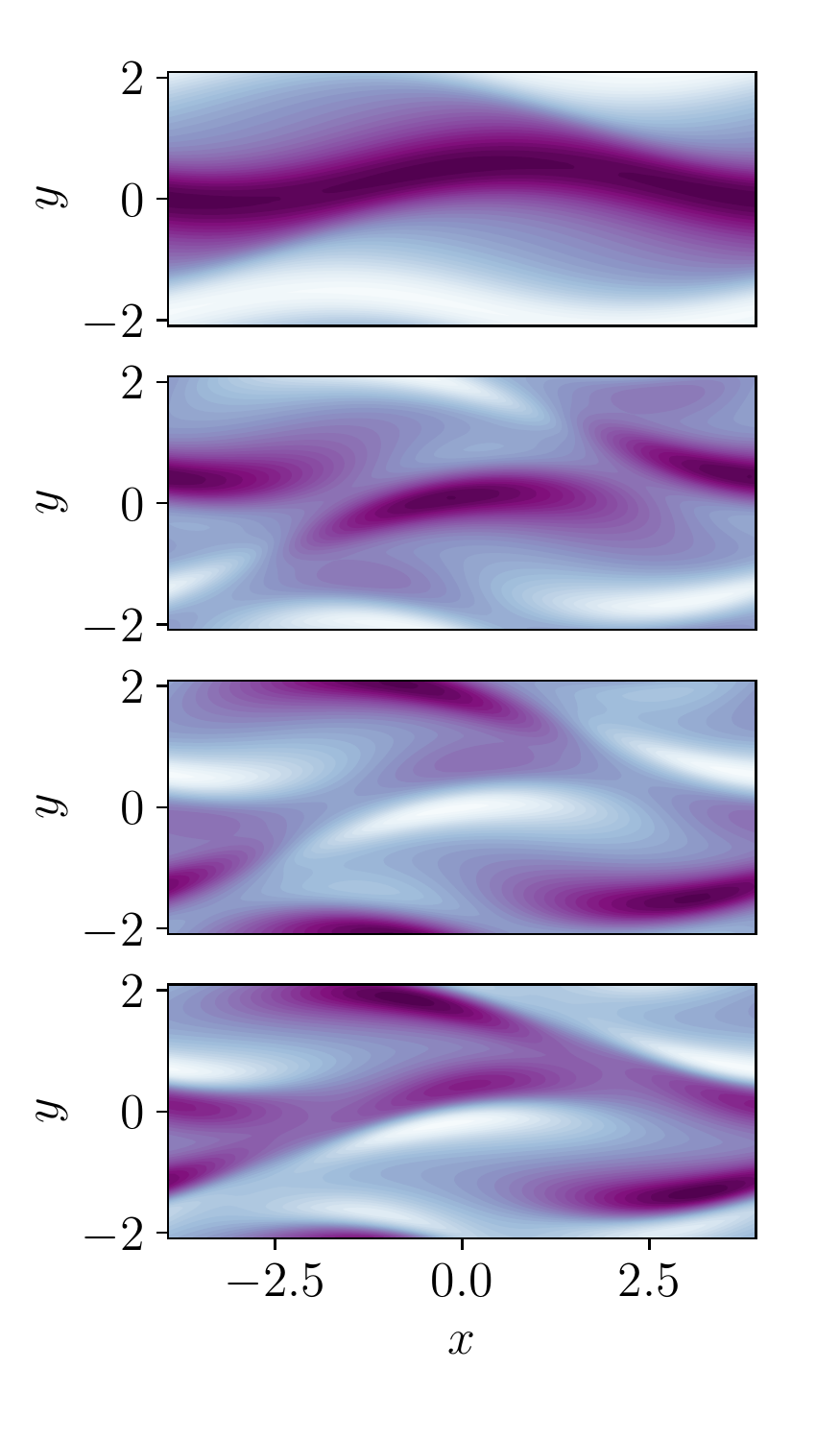}
    \includegraphics[width=0.445\textwidth]{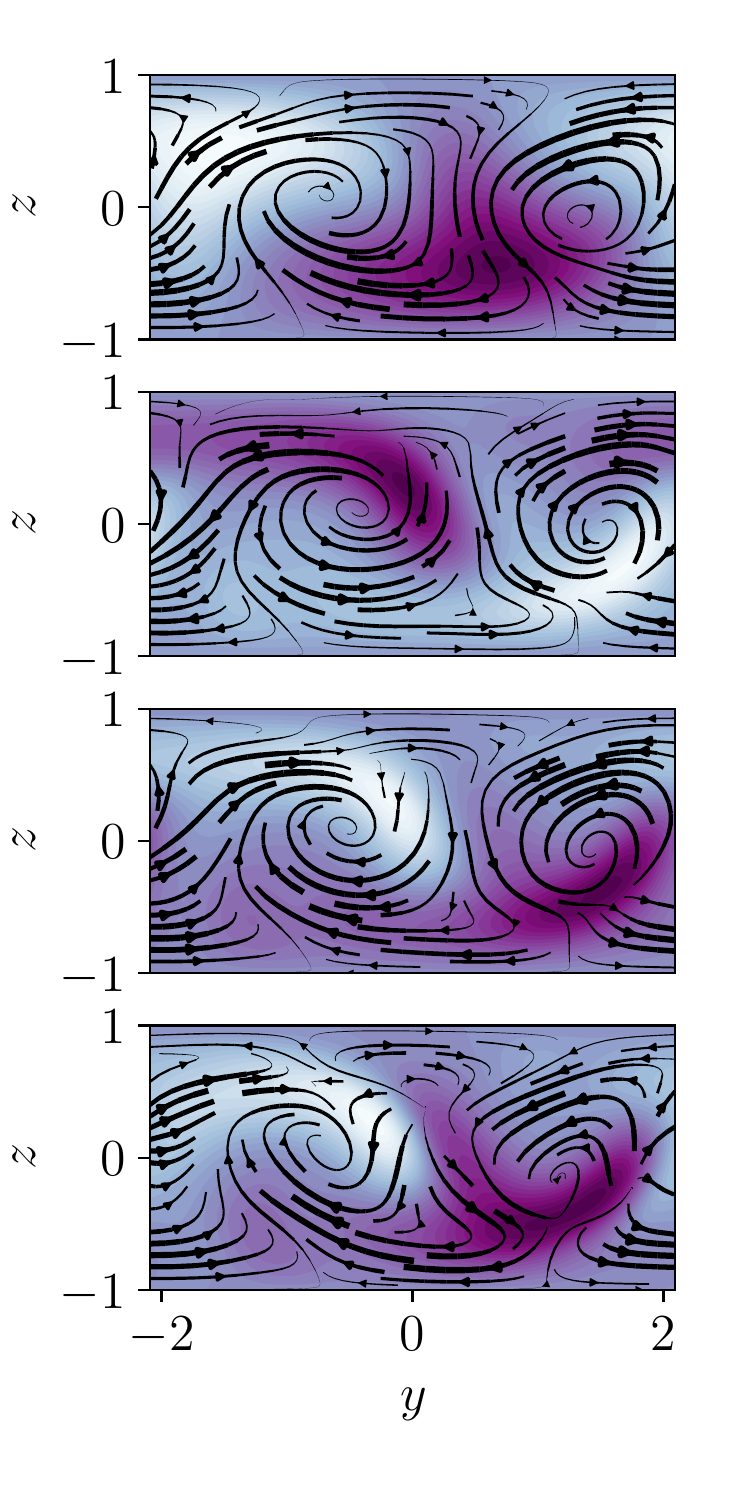}
    \caption{DMD modes corresponding to the four eigenvalues reported in figure \ref{fig:growth_eigs}, in order of increasing growth rate from top to bottom. The visualisation of each mode shows contours of the streamwise velocity at the midplane $z=0$ (left) and contours of streamwise velocity with the streamfunction overlayed on a cross-stream plane at $x=0$ (right).}
    \label{fig:growth_modes}
\end{figure}
The eigenvalues for one particular DMD calculation inside this ``growing'' region are reported in figure \ref{fig:growth_eigs}, and the corresponding DMD modes, $\{\mathbf v_n\}$, are shown in figure \ref{fig:growth_modes}. 
The neutral DMD mode is Nagata's lower branch solution, and the first growing mode is localized at the critical layer where ${\mathbf u_{LB}. \hat{\mathbf x}}=0$.  The higher order modes are qualitatively similar to the first.

%
%
\begin{figure}
    \centering
    \includegraphics[width=0.55\textwidth]{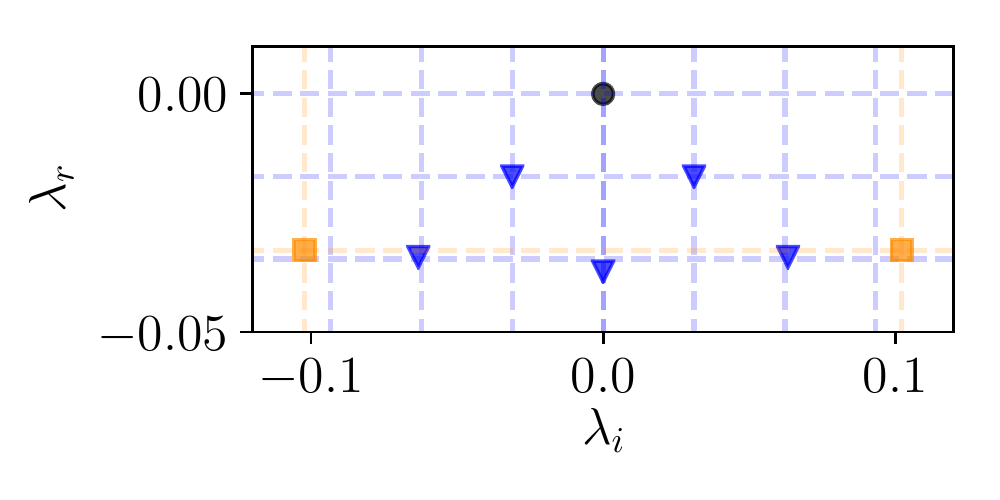}
    \caption{Eigenvalues obtained from a DMD calculation on $t\in[1000, 1100]$ from the trajectory shown in figure \ref{fig:Re135_It} (cf figure \ref{fig:dmd_window}). 
    The number of snapshots is $M=25$ and $\delta t=2$. The dashed blue lines identify integer multiples of the slowest-decaying mode, $\lambda_1\approx -0.017+0.031 \zi$.}
    \label{fig:decay_eigs}
    \includegraphics[width=0.52\textwidth]{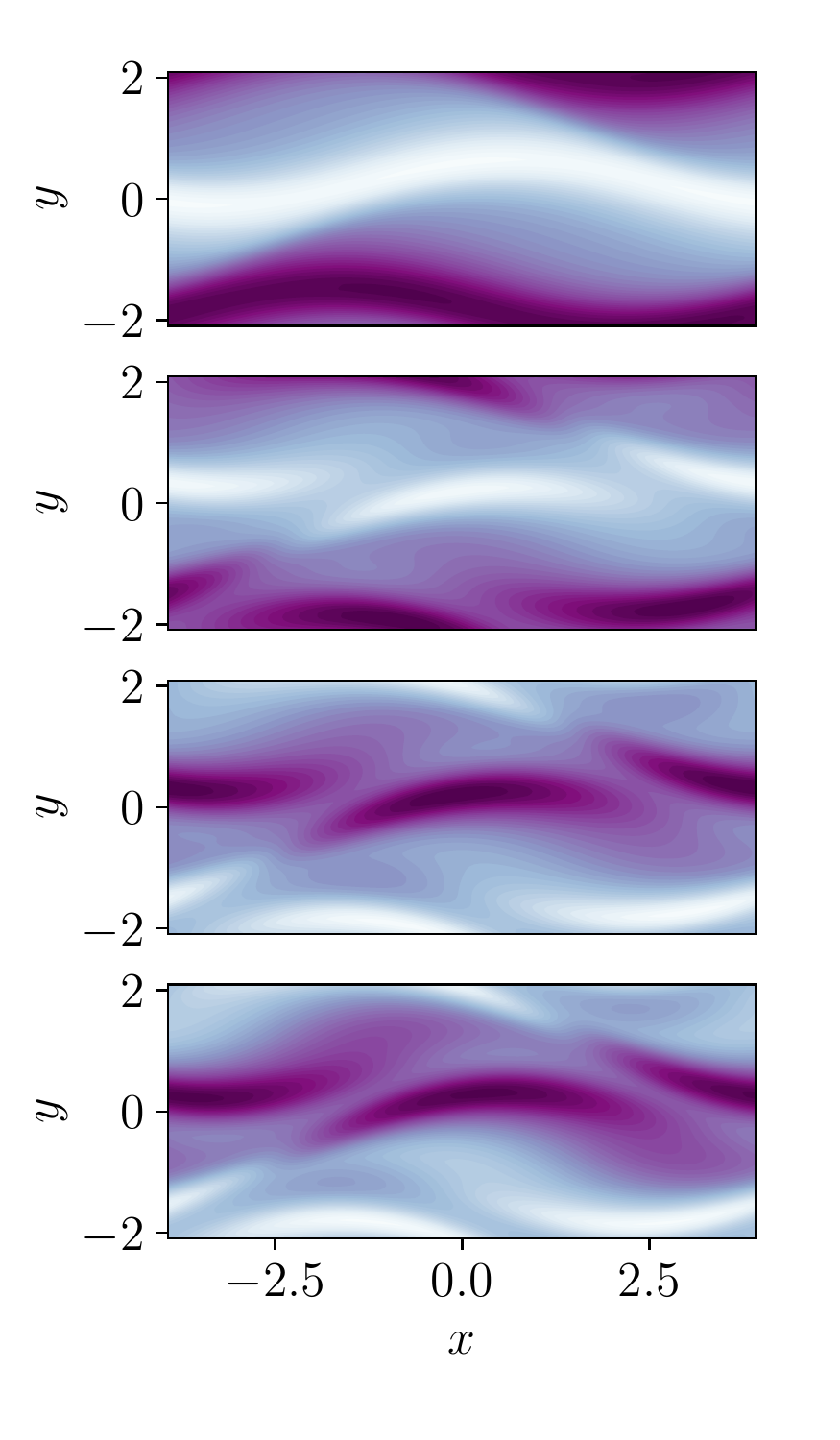}
    \includegraphics[width=0.445\textwidth]{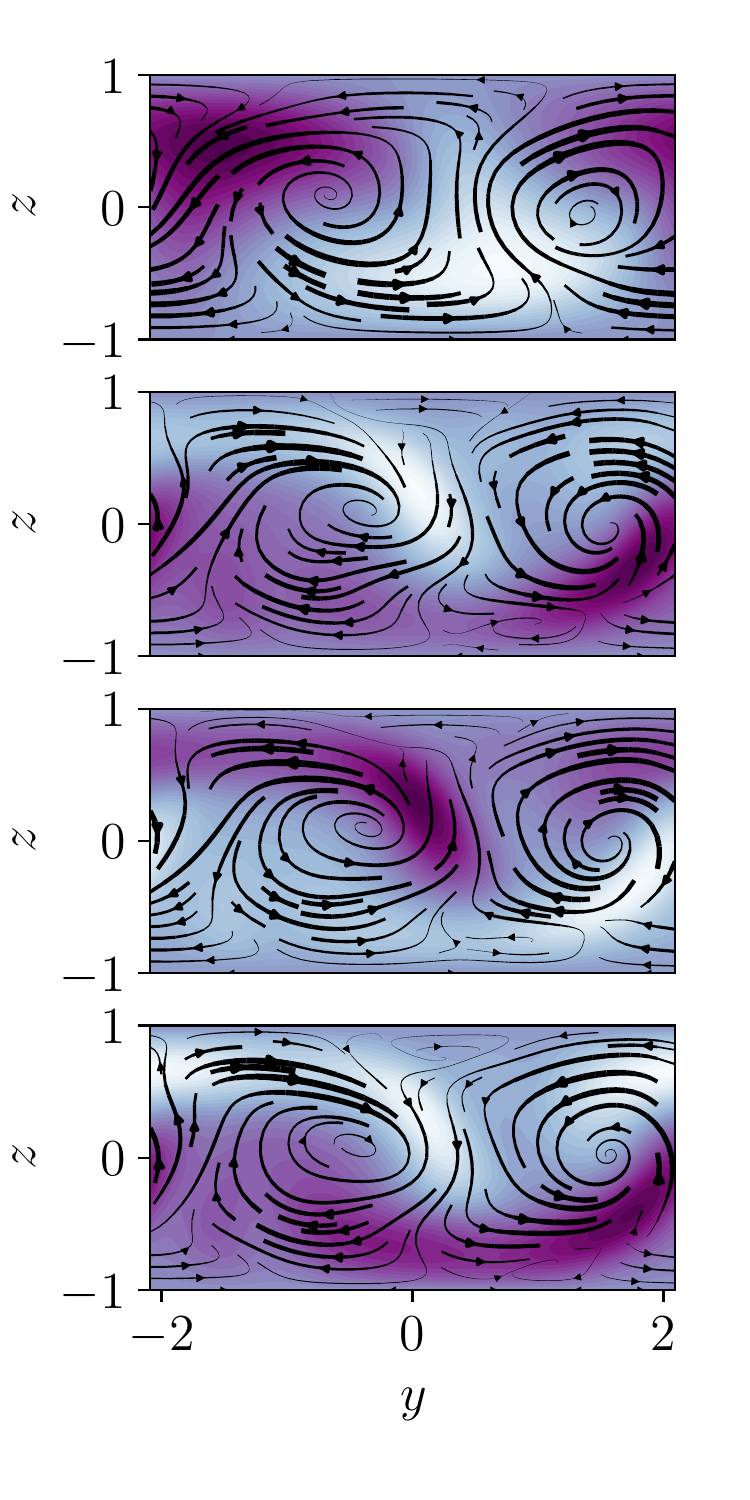}
    \caption{DMD modes (real part shown) corresponding to the four blue eigenvalues with $\lambda_i\geq 0$ reported in figure \ref{fig:decay_eigs}, in order of increasing $|\lambda|$ from top to bottom. The visualisation of each mode shows contours of the streamwise velocity at the midplane $z=0$ (left) and contours of streamwise velocity with the streamfunction overlayed on a cross-stream plane at $x=0$ (right).}
    \label{fig:decay_modes}
\end{figure}
As the DMD window is pushed further along the heteroclinic connection, pairs of (unstable) complex-conjugate eigenvalues emerge (beyond $t_F\sim 700$ in figure \ref{fig:dmd_window}) with growth rates/frequencies that are inconsistent from calculation to calculation.
However, beyond $t_F\sim 800$ a new picture emerges, and DMD identifies a variety of decaying modes that are consistent over many time windows. 
This behaviour is analogous to the crossover to the ``attracting'' expansion observed in the Stuart-Landau equation in \S2.
Furthermore, the fact that the time interval where the DMD output is inconsistent (highlighted in purple in figure \ref{fig:dmd_window}) is roughly equal to the length of the DMD time window itself, $T_w=100$, hints that there may also be a single crossover point between the two decompositions identified in the DMD rather than a finite patch of state space where neither expansion holds.
At late times the DMD identifies a single complex-conjugate pair of decaying modes in addition to a neutral eigenvalue, which indicates that trajectories spiral into the upper branch.

An example eigenvalue spectrum from the ``decaying'' region of the heteroclinic connection is reported in figure \ref{fig:decay_eigs}. 
There is a neutral mode which is the upper branch (stable) equilibrium.
The modes highlighted in blue also include the complex-conjugate pair of modes commented on above, $\lambda_1^{\pm} \approx -0.017 \pm 0.031 \zi$, 
as well as other eigenvalues built from linear combinations of this pair, i.e. $\lambda_1^+ + \lambda_1^-$, $2\lambda_1^+$ and $2\lambda_1^-$.
If the Koopman eigenfunctions associated with the least decaying pair are $\varphi_{\lambda_1}^{\pm}(\mathbf u)$, then the eigenfunctions corresponding to the higher-order modes are $\varphi_{\lambda_1}^+(\mathbf u)\varphi_{\lambda_1}^-(\mathbf u)$, $(\varphi_{\lambda_1}^+(\mathbf u))^2$ and $(\varphi_{\lambda_1}^-(\mathbf u))^2$ respectively.
The corresponding DMD modes are reported in figure \ref{fig:decay_modes}.

In addition to the family of Koopman eigenfunctions linked to the least-damped linear behaviour, there is an additional complex-conjugate pair of eigenvalues $\zeta_1^{\pm} \approx 0.03 \pm 0.10\zi $ highlighted in orange in figure \ref{fig:decay_eigs}. 
This pair of modes is consistent across the DMD calculations reported in figure \ref{fig:dmd_window}, although it is a little difficult to distinguish in that figure due to the closeness of the decay rate to $2\lambda_1$. 
These eigenvalues also describe a decaying spiral and indicate that the stable subspace around the Nagata upper branch solution is actually four-dimensional.   Their importance as the Reynolds number is increased is discussed in more detail below.

%
%
\begin{figure}
    \centering
    \includegraphics[width=0.8\textwidth]{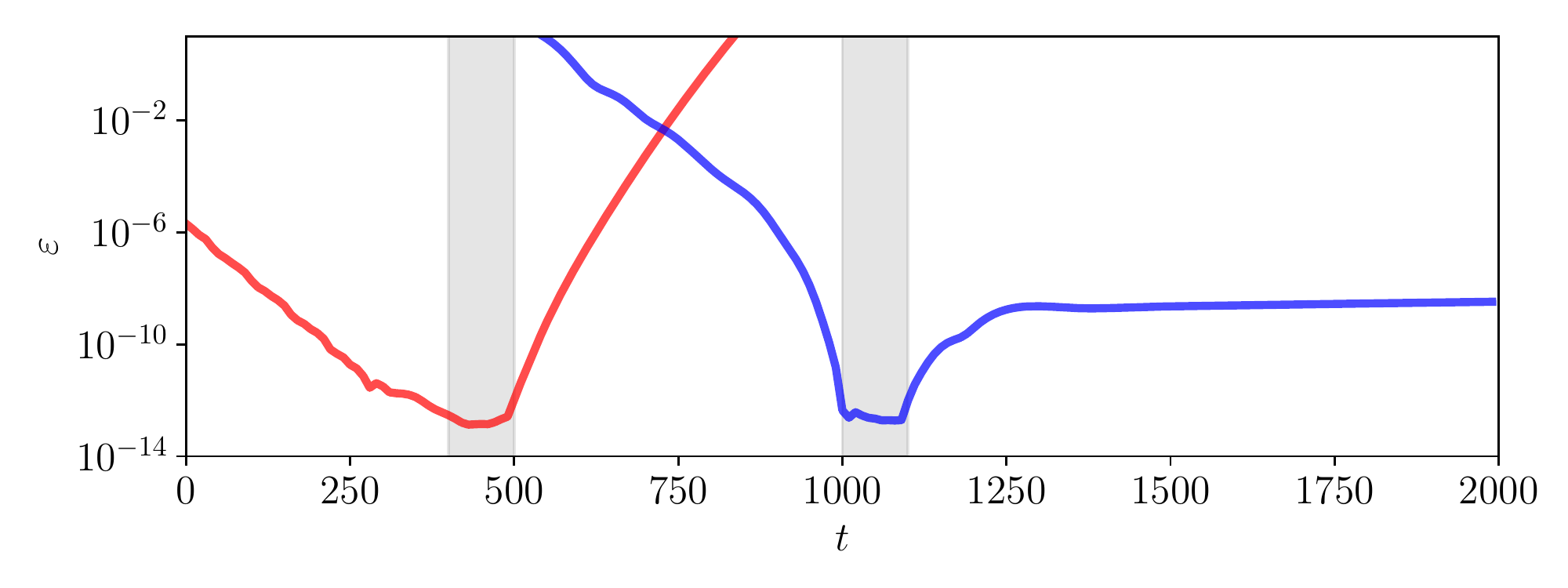}
    \caption{Error in the DMD approximation(s) (equation \ref{eqn:dmd_approx}) versus the true evolution, $\varepsilon := \|\mathbf u_{D} - \mathbf u\|/\|\mathbf u\|$. Red and blue lines identify approximations to the attracting and repelling expansions respectively, the grey regions identify windows where the DMD calculations and fitting were performed.}
    \label{fig:dmd_errs}
\end{figure}
The existence of a crossover point between the two Koopman decompositions can be explored further by using the output of the DMD calculations to construct approximations to the true trajectory.
To that end, we use the two DMD calculations reported in figures \ref{fig:growth_eigs} and \ref{fig:decay_eigs} to construct approximations to the true heteroclinic connection.
We seek a low-dimensional representation of the flow from DMD, $\mathbf u_D$, by summing over a subset $\mathcal V_{\pm}$ of the DMD modes from either the repelling or attracting regions,
\begin{equation}
    \mathbf u_D(\mathbf x,t)  =  u_C(\mathbf x) + \sum_{\lambda_j\in\mathcal V} a_j \mathbf v_j(\mathbf x) e^{\lambda_j t}.
    \label{eqn:dmd_approx}
\end{equation}
For each expansion, the modes in $\mathcal V_{\pm}$ are exactly those reported in figure \ref{fig:growth_eigs} and \ref{fig:decay_eigs}.
For the repelling expansion, this includes the neutral mode and the three unstable eigenvalues.
For the attracting expansion, the neutral mode, the five stable (blue) modes associated with the slowest-decaying spiral and the complex-conjugate (green) pair of modes associated with the second, more rapidly decaying spiral are included.

The amplitudes, $\{a_j\}$, assigned to the DMD modes are determined by a least-squares fit to the true trajectory within the DMD time window. Taking $M$ equally spaced snapshots along the fitting window separated by a time $\delta t$, the function to be minimised is
\begin{equation}
    J(\mathbf a) := \frac{1}{M} \sum_{m=0}^{M-1} \big|\boldsymbol \psi(\mathbf u(\mathbf x,m\delta t)) - \sum_j a_j\mathbf v_j e^{m\lambda_j\delta t}\big|^2.
\end{equation}
The solution to the least-squares problem for $\mathbf a$ is then
\begin{equation}
    \hat{\mathbf a} = \bigg(\sum_m (\boldsymbol \Lambda^H)^m \mathbf V^H \mathbf V\boldsymbol \Lambda^m\bigg)^{-1}\sum_m (\boldsymbol \Lambda^H)^m \mathbf V^H \boldsymbol \psi(\mathbf u(\mathbf x,m\delta t)),
\end{equation}
where $\boldsymbol \Lambda$ is a diagonal matrix where the $i^{th}$ entry is $e^{\lambda_i \delta t}$ and $\mathbf V$ is a matrix whose $i^{th}$ column is the $i^{th}$ DMD mode. 

The error betweeen the repelling and attracting approximations and the true solution are reported in figure \ref{fig:dmd_errs}. 
Unsurprisingly, the error in each case is smallest in the fitting windows themselves. 
The error also remains vanishingly small as each expansion is pushed towards its respective equilibrium solution, but both expansions blow up as they are pushed beyond an apparent crossover point at $t\approx 750$.

%
%
\begin{figure}
    \centering
    \includegraphics[width=0.48\textwidth]{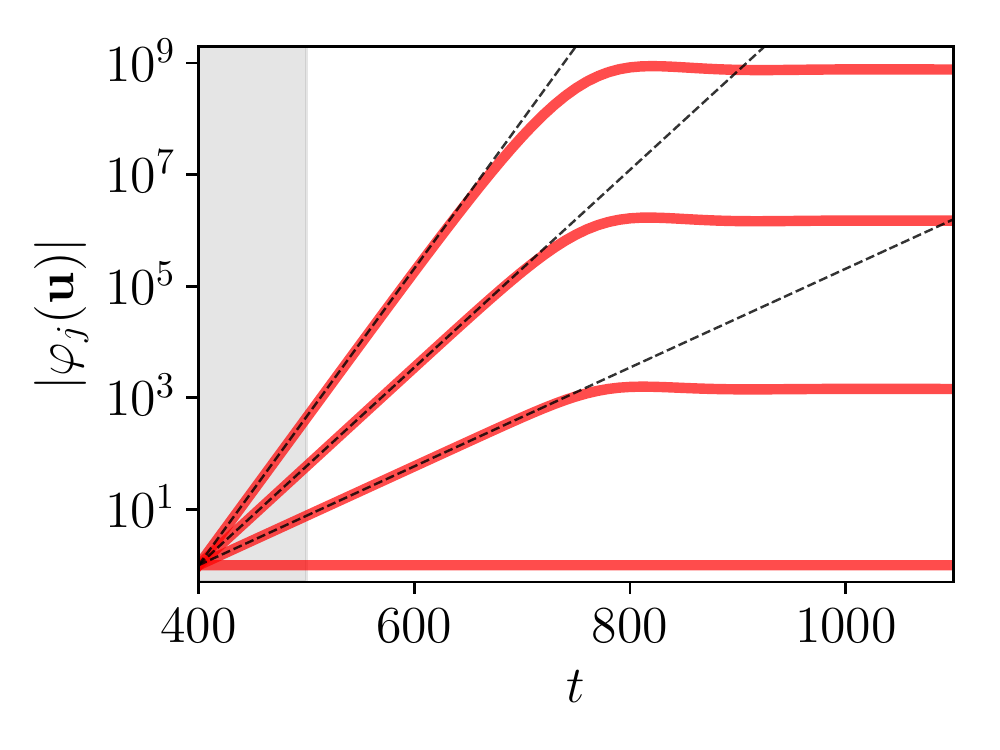}
    \includegraphics[width=0.48\textwidth]{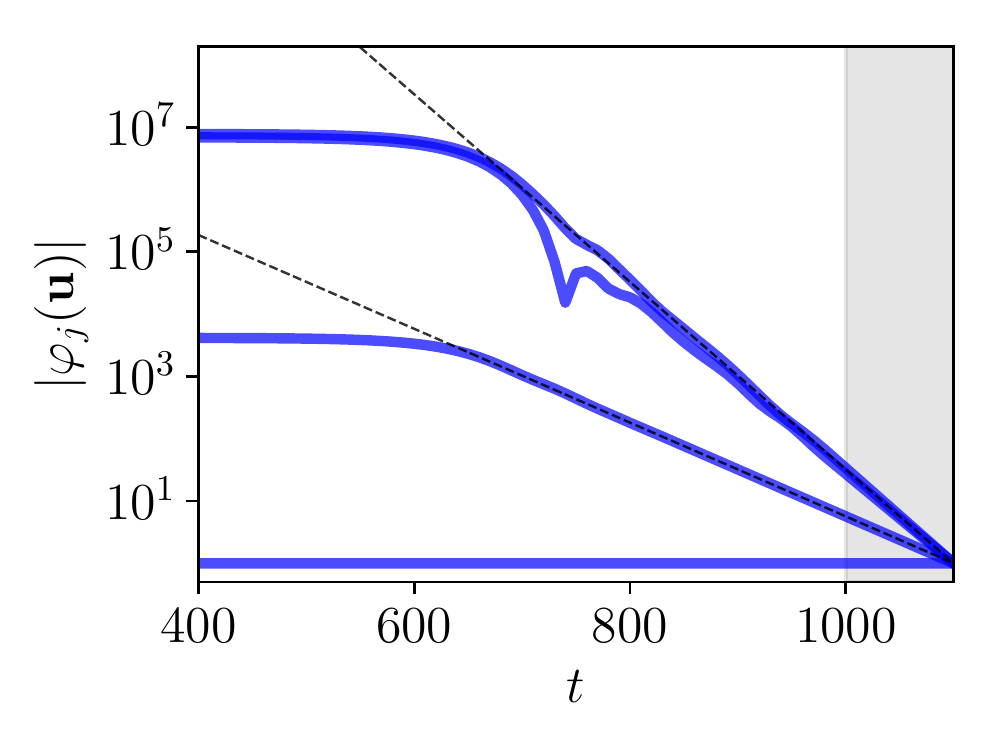}
    \caption{DMD approximations to Koopman eigenfunctions, including their time evolution, obtained in both the repelling (left - also see figures \ref{fig:growth_eigs} and \ref{fig:growth_modes}) and attracting (right - also see figures \ref{fig:decay_eigs} and \ref{fig:decay_modes}) regions. The DMD time window is highlighted in grey, and the dashed lines identify temporal behaviour $\sim e^{\lambda_j t}$. Note all repelling eigenfunctions are normalised to unity at $t=400$; the attracting eigenfunctions are normalised to unity at $t=1100$.}
    \label{fig:nagata_efns}
\end{figure}
In addition to computing the error against the true trajectory, an alternative way of assessing the output of the DMD calculations in connection to the Koopman operator is to examine the numerical approximation to the Koopman eigenfunctions.
As described in \S2, these objects are obtained from the left-eigenvectors of the DMD operator, $\{\mathbf w_j\}$ as follows \citep{Rowley2017},
\begin{equation}
    \varphi_j(\mathbf u) = \mathbf w_j^H \boldsymbol \psi(\mathbf u).
\end{equation}
The ``performance'' of DMD can be examined by evaluating this inner product for points on the trajectory beyond the DMD time window, which we do in figure \ref{fig:nagata_efns}.
It is clear that the DMD calculations on the relatively short time window $T_w=100$ have been able to accurately build locally valid representations of the Koopman eigenfunctions which remain reasonably accurate for two to three hundred advective time units beyond the observation window.
These local approximations become increasingly poor around the ``crossover point'' inferred from earlier figures (e.g. \ref{fig:dmd_window} and \ref{fig:dmd_errs}), a behaviour which again is analogous to DMD of the Stuart-Landau equation (e.g. figure \ref{fig:dmd_eigfs}).

\subsection{Higher Reynolds numbers}

In both the one dimensional Stuart-Landau equation (\S2) and the example discussed above at $Re=135$, there are only two fixed points: repelling and attracting equilibria. 
This results in a pair of Koopman expansions that extend beyond the respective repelling/attracting linear subspaces to a crossover point in state space.
Here, we increase the Reynolds number in the Nagata box to examine the consequences for Koopman decompositions and DMD when the structure of state space is complicated by the presence of additional invariant sets.
The motivation here is to explore the possibility of applying DMD to \emph{turbulent} trajectories as a method of locating nearby coherent structures and their associated Koopman mode expansions.

%
%
\begin{figure}
    \centering
    \includegraphics[width=0.48\textwidth]{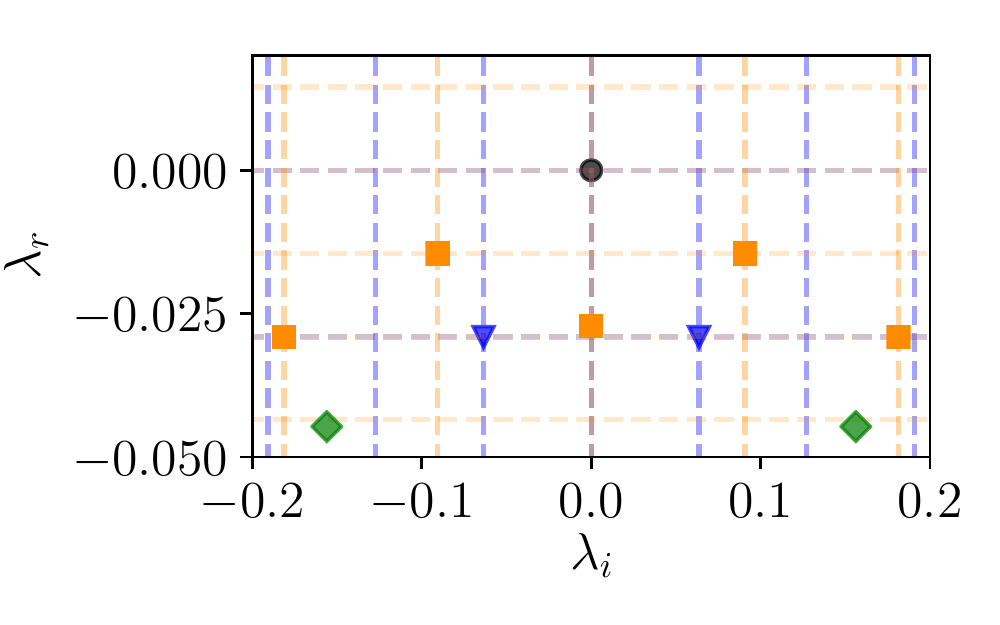}
    \includegraphics[width=0.48\textwidth]{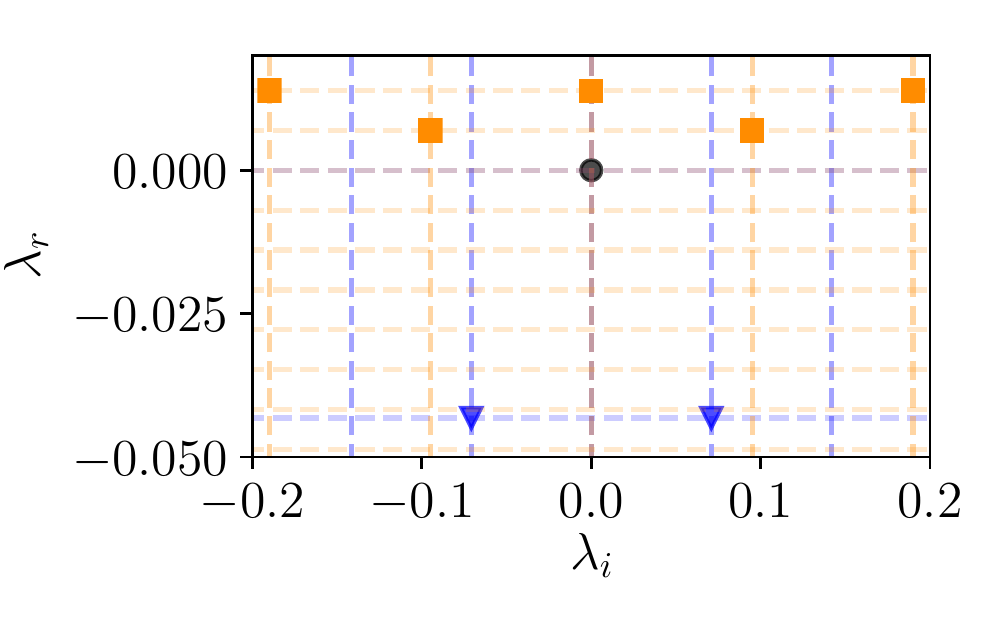}
    \caption{DMD eigenvalue spectra obtained in the vicinity of $\mathbf u_{UB}$ at $Re=140$ (left) and $Re=150$ (right). At $Re=140$, the upper branch solution is stable and the spectrum was obtained in a similar manner to that shown in figure \ref{fig:decay_eigs} at $Re=135$ ($M=25$ snapshots in a timewindow of length $T_w=100$, with $\delta t=2$). At $Re=150$, the DMD was performed on a trajectory  $\mathbf f^t(\tilde{\mathbf u}_{UB})$ over $t\in [0,700)$, where $\tilde{\mathbf u}_{UB}$ is the numerical approximation to the upper branch equilibrium converged using Newton-GMRES over a time interval $T=4$. $M=25$ snapshots were used with spacing $\delta t=1$.}
    \label{fig:nagata_Re140_150_eigs}
\end{figure}
The eigenvalue spectra obtained in figures \ref{fig:dmd_window} and \ref{fig:decay_eigs} indicate that the stable subspace 
around the upper branch is four-dimensional (two orthogonal spirals).
The DMD also revealed the higher-order Koopman eigenvalues required to propagate $\mathbf u$ beyond the linear subspace.
The second, more rapidly decaying spiral (orange squares in figure \ref{fig:decay_eigs}) becomes increasingly dominant in the dynamics as the Reynolds number is increased. 
This behaviour is apparent in figure \ref{fig:nagata_Re140_150_eigs}, where we report DMD eigenvalues from trajectories very close to the upper branch $\mathbf u_{UB}$ at $Re\in \{140, 150\}$. 

At $Re=140$ the eigenvalues associated with the second spiral in the linear subspace around $\mathbf u_{UB}$ ($\lambda = \zeta_1^{\pm} \approx = -0.015 \pm 0.091\zi$;  orange in figure \ref{fig:nagata_Re140_150_eigs}), which had only a weak effect on the dynamics at $Re=135$, have become the most slowly decaying to dominate the linearized dynamics. 
In addition, higher order Koopman eigenvalues in the same family as $\zeta_1^{\pm}$ are also obtained from the DMD (e.g. $2\zeta_1$ associated with $\varphi_{\zeta_1}^2(\mathbf u)$) and are highlighted with dashed lines.  
The decay rate of the first spiral (blue triangles) has approximately doubled.
Note that, in addition to the two families of Koopman eigenvalues associated with the dynamics in the linear subspace, there are also eigenvalues which are connected to products of Koopman eigenfunctions from these families.
For example, in figure \ref{fig:nagata_Re140_150_eigs} the green diamonds identify eigenvalues $\lambda_1^{\pm} + \zeta_1^{\pm}$ associated with eigenfunctions $\varphi^{\pm}_{\lambda_1}(\mathbf u)\varphi^{\pm}_{\zeta_1}(\mathbf u)$.

At around $Re\approx 145$ the complex conjugate pair of eigenvalues associated with the dominant spiral, $\zeta^{\pm}_1$, cross the imaginary axis (not shown) to become unstable, and
a stable period orbit (SPO) is born in a Hopf bifurcation off $\mathbf u_{UB}$.
This behaviour is apparent in the results of DMD at $Re=150$ in figure \ref{fig:nagata_Re140_150_eigs}.
In addition to the unstable pair of eigenvalues associated with the dynamics in the linear subspace, $\zeta^{\pm}_1 \approx 0.007 \pm 0.095 \zi$, higher order Koopman eigenvalues are again observed, and correspond to products of the Koopman eigenfunctions $\varphi^{\pm}_{\zeta_1}(\mathbf u)$. 

The destabilisation of the upper branch solution and the emergence of a SPO has further consequences for DMD.
There are now three crossover points associated with Koopman decompositions around each of the three invariant sets ($\mathbf u_{LB}$, $\mathbf u_{UB}$ and the SPO), and DMD will only ``work'' if it is restricted to a particular expansion zone. 
The results highlight the care that must be taken in more complex flows with many exact coherent states buried in the turbulent attractor.

%
%
\begin{figure}
    \centering
    \includegraphics[width=0.48\textwidth]{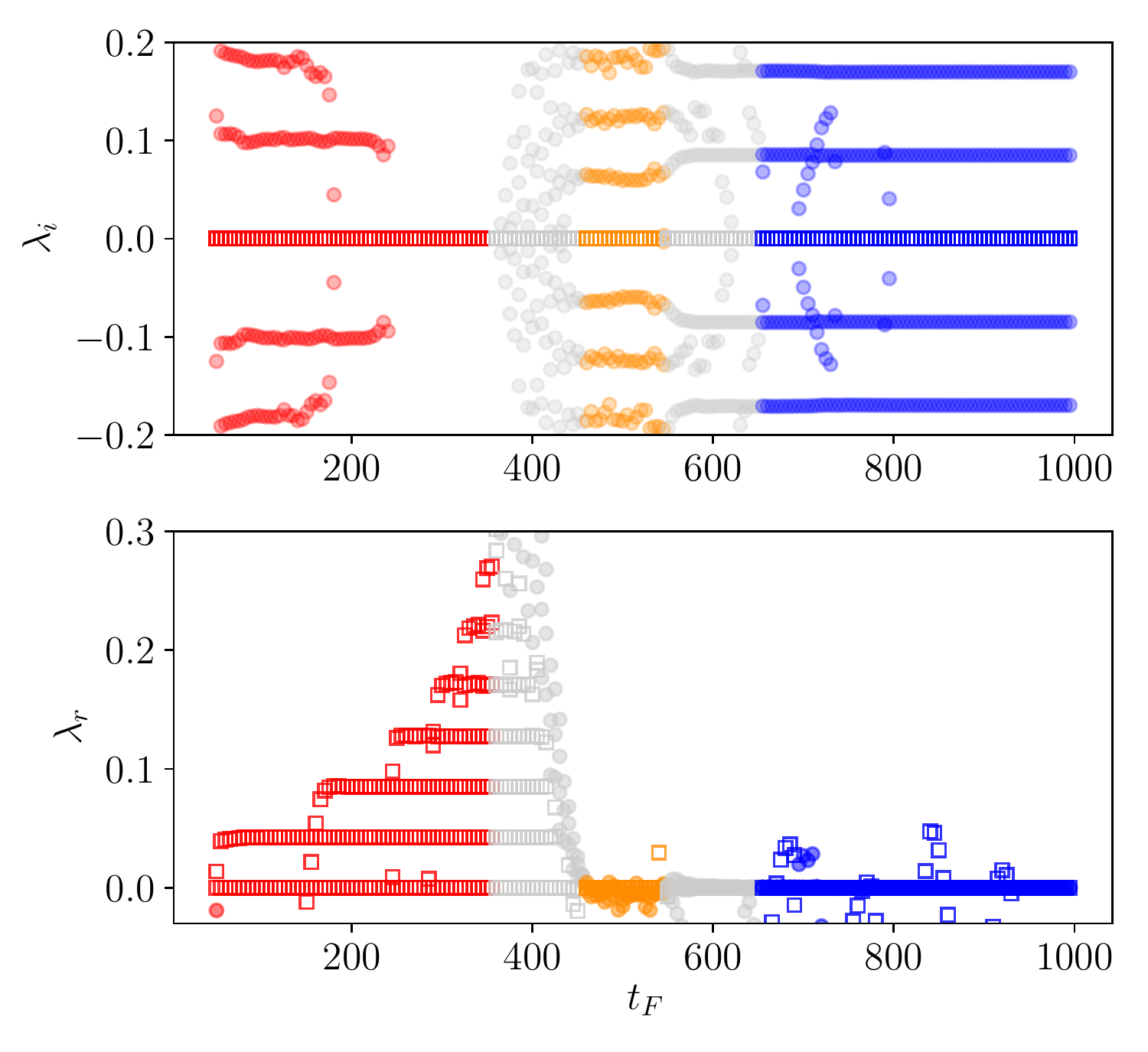}
    \includegraphics[width=0.48\textwidth]{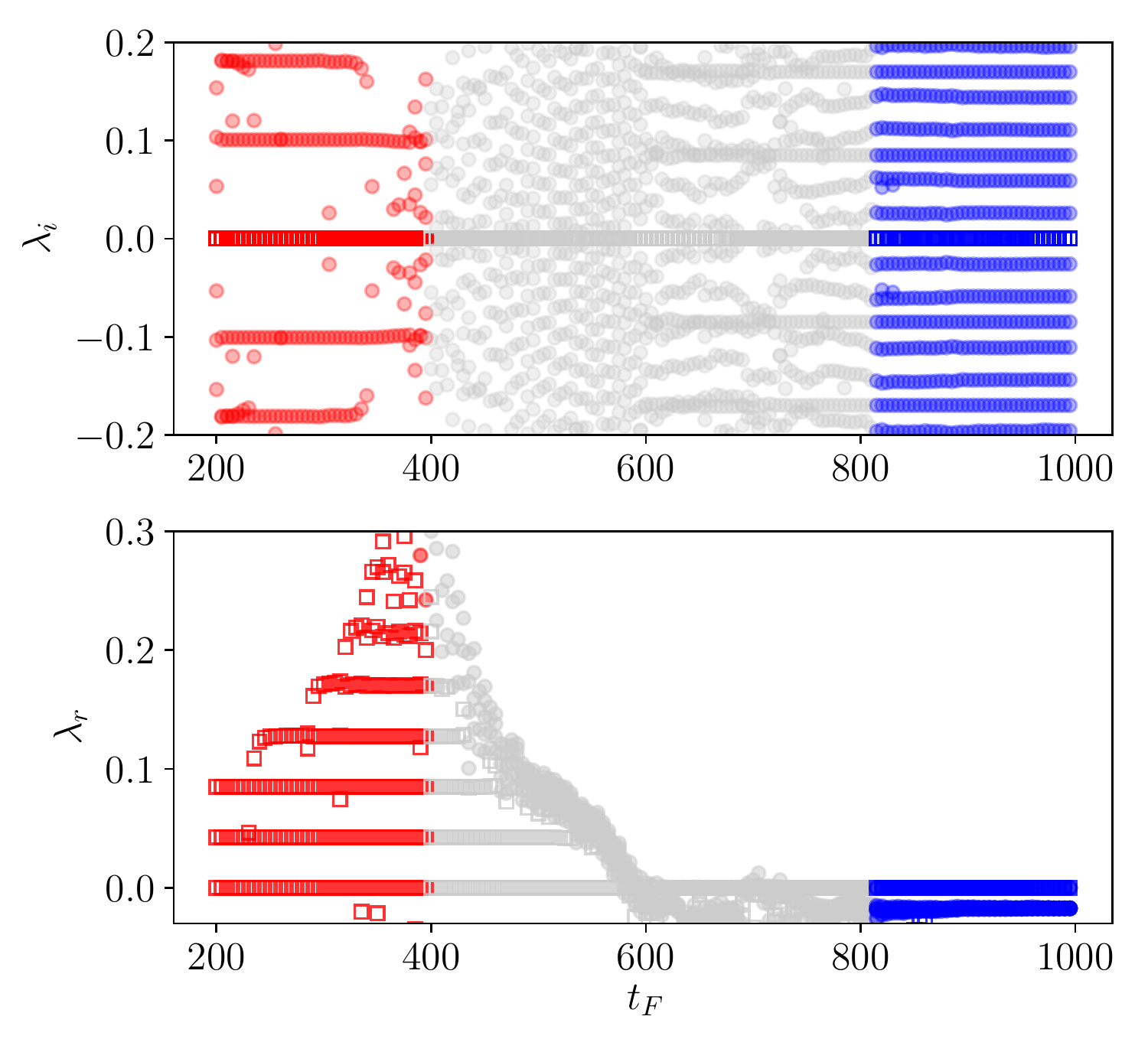}
    \caption{(Top) Real and (bottom) imaginary components of DMD eigenvalues obtained on shorter time windows passed through a trajectory running from the linear subspace of $\mathbf u_{LB}$ to the SPO at $Re=150$ (the variable $t_F$ is the ``final'' time of each time window). Left: time windows of length $T_w=50$ are used with $M=50$ snapshots. Right: $T_w=200$ with $M=150$ snapshots. Snapshot spacing is $\delta t = 1$ for both sets of calculations. The colouring serves as a guide for the eye, with red, orange  and blue identifying expansions around $\mathbf u_{LB}$, $\mathbf u_{UB}$ and the SPO respectively.}
    \label{fig:dmd_Re150}
\end{figure}
\begin{figure}
    \centering
    \includegraphics[width=0.48\textwidth]{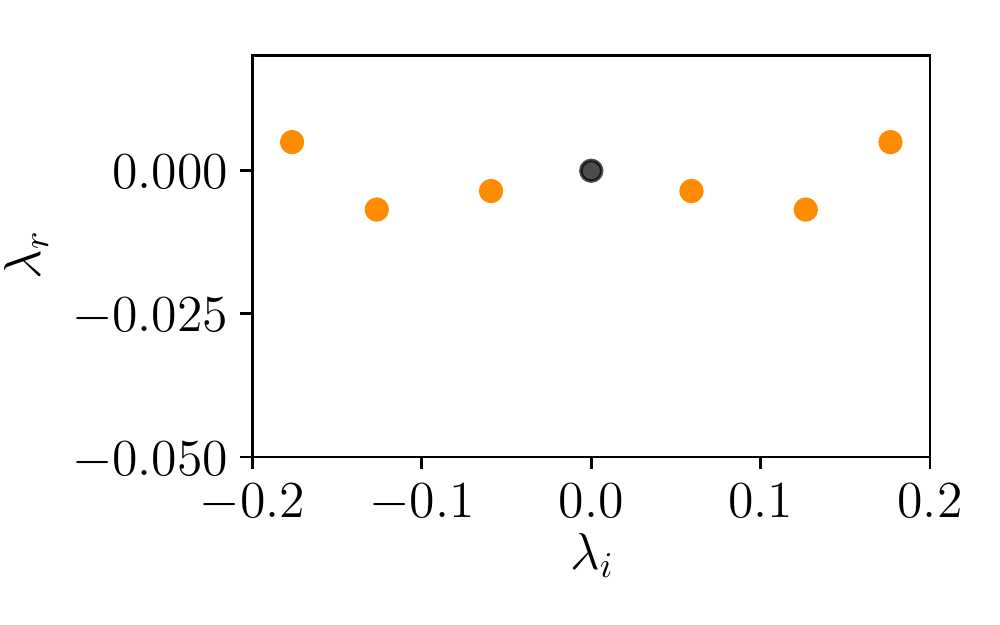}
    \includegraphics[width=0.48\textwidth]{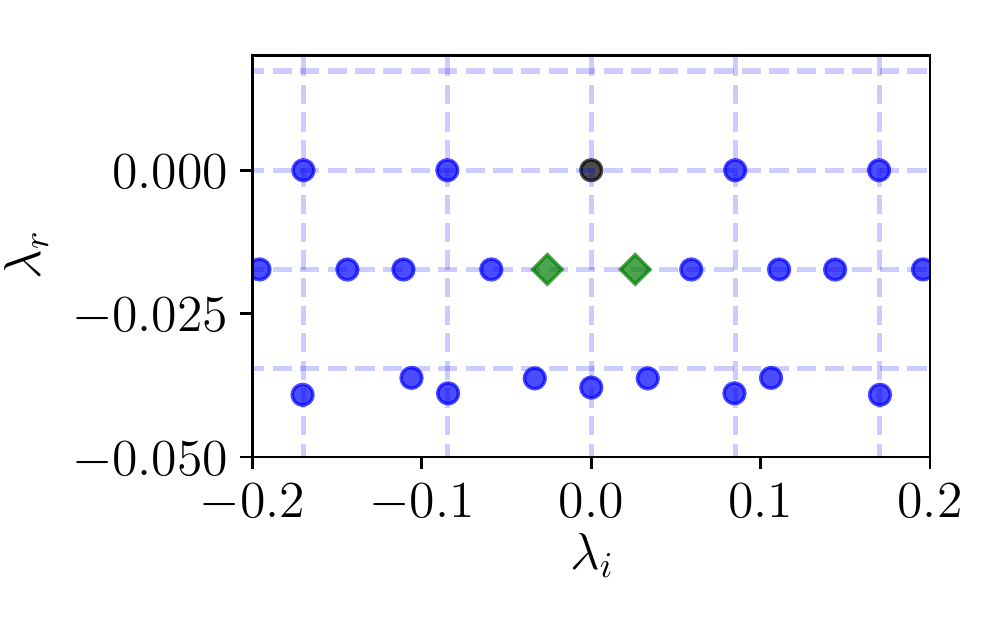}
    \caption{DMD eigenvalues obtained from the $\mathbf u_{LB}\to$ SPO heteroclinic connection at $Re=150$. Left: time window $t \in [450, 500]$, with $M=50$ snapshots, $\delta t=1$ (c.f. upper branch spectrum in \ref{fig:nagata_Re140_150_eigs}). Right: time window $t\in [650,1000]$, with $M=200$ snapshots, $\delta t=1$. The modes highlighted in green are the Floquet multipliers $\mu \approx -0.017\pm 0.026\zi$ (from the form $e^{\mu \delta t}$). Dashed lines identify harmonics of the fundamental frequency of the periodic orbit.}
    \label{fig:Re150_spectra}
\end{figure}
To demonstrate the restrictions placed on DMD, we consider again a trajectory beginning in the linear subspace around $\mathbf u_{LB}$, which now collapses into the SPO as $t\to \infty$.
Similar to our approach for the $\mathbf u_{LB}\to \mathbf u_{UB}$ connection at $Re=135$, we perform many DMDs in a fixed time window which is passed along the finite time approximation to the heteroclinic connection.
The results of these calculations for two DMD time windows, $T_w \in \{50, 200\}$, are reported in figure \ref{fig:dmd_Re150}.

For the shorter DMD time window, $T_w=50$, three distinct trends in the eigenvalues are observed.
As the trajectory moves away from the lower branch solution, the DMD locates the positive, real eigenvalue associated with the growth rate along the single unstable direction, before identifying the integer multiples of this growth rate corresponding to the higher order Koopman modes. 
This behaviour is analogous to that found at $Re=135$ (see figure \ref{fig:dmd_window}); like that earlier $\mathbf u_{LB}\to\mathbf u_{UB}$ connection, there is also a breakdown in the DMD/Koopman eigenvalues, here at $t_F \sim 400$.
The region where there is inconsistency between successive DMD calculations is roughly twice the length of the DMD time window, which suggests the presence of a crossover point.
Beyond $t_F\sim 450$, there is a clear repeated frequency in the DMD eigenvalues, $\lambda_i\approx 0.064$, consistent over roughly 100 advective time units. 
This frequency is close to that associated with the stable spiral into the upper branch ($\lambda_i = 0.069$, see figure \ref{fig:nagata_Re140_150_eigs}).
However, the DMD is unable to resolve the associated decay rate correctly, or obtain the complex conjugate pair of unstable modes associated with $\mathbf u_{UB}$ at this Reynolds number.
An individual eigenvalue spectrum from this region is reported in figure \ref{fig:Re150_spectra} and should be contrasted with those obtained on trajectories starting in the linear subspace of $\mathbf u_{UB}$ (figure \ref{fig:nagata_Re140_150_eigs}). 
A pair of unstable eigenvalues are found, though their growth rate and frequency do not correspond to the unstable directions identified in figure \ref{fig:nagata_Re140_150_eigs}.
It is likely that the trajectory simply does not go close enough to the upper branch equilibrium to accurately distinguish the correct form of the neutral mode ($\mathbf u_{UB}$ itself) from the slowly growing eigenvalues, and this error contaminates the rest of the spectrum. 
Finally, at around $t_F\sim 600$ the output of the DMD calculations jumps again. 
There are clear repeated harmonics of a fundamental frequency, $\omega_f = 0.085$ ($\lambda_r$ is very close to zero), which corresponds to a periodic orbit with period $T = 73.9$. 
Occasionally, the DMD erroneously identifies growing modes, while the array of decaying eigenvalues one would expect to find around a stable limit cycle is absent \citep{Bagheri2013}. 

To accurately determine the Koopman eigenvalues around the SPO, a much longer time window is required.
For example, the longer time window considered in figure \ref{fig:dmd_Re150}, $T_w=200$, no longer shows exponentially unstable modes in the collapse onto the SPO.
Instead, there is an array of decaying eigenvalues, all with decay rate $\lambda_r \approx -0.017$. 
Each harmonic of the SPO is flanked by a pair of decaying modes, $\lambda = n\omega_f\zi + (\mu_r \pm \mu_i\zi)$, indicating the presence of a pair of stable Floquet multipliers $e^{\mu T}$, with $\mu=-0.017 \pm 0.026\zi$.
To obtain higher order Koopman modes associated with the SPO \citep[see][]{Bagheri2013}, even longer time windows are required. 
For example, the decay rate $2\mu_r$ is observed in figure \ref{fig:Re150_spectra} with a slightly longer time window $T_w=350$, although only approximately and there are eigenvalues missing. 

A consequence of the longer time window required for a more accurate resolution of the Koopman eigenvalues around the SPO is the loss of any indication of the presence of the upper branch in figure \ref{fig:dmd_Re150}. The time window is longer than the residence time in the upper branch expansion region, so all DMD calculations that see $\mathbf u_{UB}$ contain at least one crossover point. 
In addition to a large time window, $T_w>T$, DMD calculations which are able to accurately resolve the Koopman eigenvalues around the periodic orbit require many snapshot pairs.
In a turbulent flow with unstable periodic orbits (UPOs), each with many different Floquet multipliers, these requirements are unlikely to be achievable in practice.
However, the fact that DMD time windows which are shorted than the period, $T_w<T$, are still able to identify the fundamental frequencies and associated mode shapes indicates that DMD may be a useful alternative to a recurrent flow analysis in generating guesses for UPOs.

\section{Conclusion}
\label{sec:conc}
In this paper we have examined how the presence of multiple simple invariant solutions in a nonlinear dynamical system affects the construction of Koopman expansions for the state variable.
We showed how an inverse Laplace transform can be used to obtain Koopman mode decompositions if the Koopman eigenvalues are purely real, before applying this technique to the Stuart-Landau equation.
The solution revealed two possible Koopman expansions, each corresponding to a particular fixed point of the dynamical system.
There is a \emph{crossover point} in state space where one expansion breaks down and the other takes over.
The success of DMD to locate Koopman eigenvalues depends critically on the location of the crossover point: a DMD performed over a time window which contains the crossover point will fail.  

We then applied DMD to some heteroclinic connections of the Navier-Stokes equations in Couette flow at low Reynolds number.
The results confirm the existence of a different Koopman expansion around each simple invariant solution.
Again, the ability of DMD to discover these decompositions is constrained by the presence of crossover points in state space.
Only a DMD restricted to a particular ``expansion region'' can identify the underlying Koopman eigenvalues and modes. 

These findings suggest that DMD may still be a useful tool for finding exact coherent structures near to a turbulent orbit provided the data window is taken small enough so that only the neighbourhood of one coherent structure is sampled. The approach when searching for equilibria is more refined than supplying snapshots of the turbulent field and trying to converge a steady solution with GMRES-Hookstep. While for periodic orbits, the DMD time window need not contain a ``near recurrence'' to identify the relevant frequencies and mode shapes that serve as the input to a root-finding algorithm.  We hope to report our results using DMD to extract coherent structures from turbulent flows in the near future.

\section*{Appendix: Carleman linearization}

In this appendix, we show that the two Koopman expansions found in \S \ref{KoopmanEx}  emerge naturally from Carleman linearization about the respective equilibria. The 1D nonlinear equation (\ref{scaled_1D_eqn}) can be converted into an infinite dimensional linear system 
\begin{equation}
\frac{\dd x_n}{\dd t}=(2n-1)(x_n-x_{n+1})
\end{equation}
by  defining new variables $x_n:=R^{2n-1}(t)$ for $n \in {\mathbb N}$. A truncated version of this system ($x_n=0$ for $n > N$) should approximate the dynamics  in the neighbourhood of the equilibria $R=0$ where neglected variables should be negligible.
This  truncated system is
\begin{equation}
\frac{\dd \mathbf x}{\dd t}= \mathbf L \mathbf x
\label{linear}
\end{equation}
where $\mathbf x=(x_1\; x_2\; \dots \; x_N)^T$ and L$_{n\, \!n}=2n-1$, L$_{n\, \!n+1}=-(2n-1)$ with L$_{n\, \! m}=0$ otherwise.
The matrix $\mathbf L$ is (upper) triangular so its eigenvalues can be read off from the diagonal and correspond to the first $N$ Koopman eigenvalues  relevant for an expansion around $r=0$. The {\em left}                
 eigenvector ${\mathbf w}(n)$ of the eigenvalue $2n-1$  corresponds to the Koopman eigenfunction
\begin{equation}
\phi_{2n-1}(R)=
\left( \frac{R^2}{1-R^2} \right)^{\tfrac{(2n-1)}{2}}=
    R^{2n-1} \biggl[ 1+ \frac{2n-1}{2}R^2+\frac{(2n-1)(2n+1)}{2^2 \,2!}R^4+\cdots \biggr].
\end{equation}
where the expansion in $R^2$ is truncated at $R^{2N-1}$ (i.e. $w_j(n)=0$ for $j < n$, $w_n(n)=1$, $w_{n+1}(n)=(2n-1)/2$ etc.).
In contrast, the {\em right} eigenvector ${\mathbf v}(n)$ is exactly the Koopman mode for the observable vector $\mathbf x$ since this naturally truncates at the $n^{th}$ component. To see this, recognise that the $p^{th}$ observable is 
\begin{equation}
x_p(t):=R^{2p-1}(t)=\frac{ R_0^{2p-1} e^{(2p-1)t}}{ (1-R_0^2)^{(2p-1)/2}} \left[1+ \frac{R_0^2}{1-R_0^2} e^{2t}\right]^{-(2p-1)/2}
\end{equation}
using the exact solution (\ref{eqn:exact_soln}) and so only exponentials $e^{(2n-1)t}$ for $n \geq p$ are needed to express this evolution. In other words, the $n^{th}$ Koopman eigenfunction is only needed in an expansion for $x_p$ if $n \geq p$ so the $n^{th}$ Koopman mode will have zero  components beyond the $n^{th}$ component.

An equivalent Carleman linearization procedure can be carried out around the $R=1$ attractor using a  new variable $z:=1-R$ so that (\ref{scaled_1D_eqn}) becomes
\begin{equation}
\frac{\dd z}{\dd t}=-2z+3z^2-z^3.
\end{equation}
Defining new variables $x_n:=z^n(t)$ for $n \in {\mathbb N}$ and truncating after $x_N$ leads to the linear system (\ref{linear}) with
L$_{n\,\!n}:=-2n$, L$_{n\,\!n+1}:=3n$, L$_{n\;\!n+2}=-n$ the only non zero matrix elements. As before, the eigenvalues of L are the first $N$ Koopman eigenvalues,  the left eigenvectors represent truncated approximations of the Koopman eigenfunctions and the right eigenvectors are exactly the Koopman modes.  

Carleman linearization can also be applied centred on non-equilibria but the resulting linear system is then only affine rather than linear. Rewriting
(\ref{scaled_1D_eqn}) in favour of $z:=R-R^*$ where $R^*-R^{*3} \neq 0$ gives
\begin{equation}
\frac{\dd z}{\dd t}=(R^*-R^{*3})+(1-3R^{*2})z-3R^*z^2-z^3.
\end{equation}
With $x_n:=z^n$, this has form
\begin{equation}
\frac{\dd \mathbf x}{\dd t}= \mathbf L \mathbf x+\mathbf b
\end{equation}
where the only non-zero elements of $\mathbf L$ are
\begin{equation}
{\rm L}_{jk}:= \left\{ 
\begin{array}{ll}
(R^*-R^{*3})j &  k=j-1, \\
(1-3R^{*2}) j &  k=j,\\
-3R^*j            & k=j+1,\\
-j                   & k=j+2,
\end{array}
\right.
\end{equation}
with $\mathbf b:=[(R^*-R^{*3}) \; 0\; 0\; \ldots 0]^T$ and formal solution
\begin{equation}
{\mathbf x}(t) = e^{{\mathbf L}t} \left( \mathbf L^{-1} \mathbf b + {\mathbf x}(0) \right)-\mathbf L^{-1} \mathbf b.
\label{formal_soln}
\end{equation}
The dynamics of the truncated system provides a good approximation to the full dynamics  in the neighbourhood of $R=R^*$ but  clearly cannot be  captured by a sum of exponentially evolving components only due to the presence of the $-\mathbf L^{-1} \mathbf b$ term. The presence of this  term reflects the fact that the linearization has been performed about a non-equilibrium.

\bibliographystyle{jfm}
\bibliography{heteroclinic_1}

\begin{thebibliography}{48}
\expandafter\ifx\csname natexlab\endcsname\relax\def\natexlab#1{#1}\fi
\def\au#1{#1} \def\ed#1{#1} \def\yr#1{#1}\def\at#1{#1}\def\jt#1{\textit{#1}}
  \def\bt#1{#1}\def\bvol#1{\textbf{#1}} \def\vol#1{#1} \def\pg#1{#1}
  \def\publ#1{#1}\def\arxiv#1{#1}\def\org#1{#1}\def\st#1{\textit{#1}}

\bibitem[Arbabi \& Mezi\'c(2017)]{Arbabi2017}
{\sc \au{Arbabi, H.} \& \au{Mezi\'c, I.}} \yr{2017}  \at{{Study of dynamics in
  post-transient flows using Koopman mode decomposition}}.  \jt{Phys. Rev.
  Fluids}  \bvol{2},  \pg{124402}.

\bibitem[Avila {\em et~al.\/}(2013)Avila, Mellibovsky, Roland \&
  Hof]{Avila2013}
{\sc \au{Avila, M.}, \au{Mellibovsky, F.}, \au{Roland, N.} \& \au{Hof, B.}}
  \yr{2013}  \at{{Streamwise-localized solutions at the onset of turbulence in
  pipe flow}}.  \jt{Physical Review Letters}  \bvol{110},  \pg{224502}.

\bibitem[Bagheri(2013)]{Bagheri2013}
{\sc \au{Bagheri, S.}} \yr{2013}  \at{{Koopman-mode decomposition of the
  cylinder wake}}.  \jt{J. Fluid Mech.}  \bvol{726},  \pg{596--623}.

\bibitem[Brand \& Gibson(2014)]{BrandGibson2014}
{\sc \au{Brand, E.} \& \au{Gibson, J.~F.}} \yr{2014}  \at{{ A doubly localized
  equilibrium solution of plane Couette flow}}.  \jt{Journal of Fluid
  Mechanics}  \bvol{750},  \pg{R3}.

\bibitem[Brunton {\em et~al.\/}(2016{\natexlab{{\em a\/}}})Brunton, Johnson,
  Ojemann \& Kutz]{BruntonB2016}
{\sc \au{Brunton, B.~W.}, \au{Johnson, L.~A.}, \au{Ojemann, J.~G.} \& \au{Kutz,
  J.~N.}} \yr{2016{\natexlab{{\em a\/}}}}  \at{Extracting spatial–temporal
  coherent patterns in large-scale neural recordings using dynamic mode
  decomposition}.  \jt{Journal of Neuroscience Methods}  \bvol{258},
  \pg{1--15}.

\bibitem[Brunton {\em et~al.\/}(2016{\natexlab{{\em b\/}}})Brunton, Brunton,
  Proctor \& Kutz]{Brunton2016}
{\sc \au{Brunton, S.~L.}, \au{Brunton, B.~W.}, \au{Proctor, J.~L.} \& \au{Kutz,
  J.~N.}} \yr{2016{\natexlab{{\em b\/}}}}  \at{{Koopman invariant subspaces and
  finite linear repesentations of nonlinear dynamical systems for control}}.
  \jt{PLoS ONE}  \bvol{11}~(2).

\bibitem[Carleman(1932)]{Carleman1932}
{\sc \au{Carleman, T.}} \yr{1932}  \at{{Application de la theories des
  equations integrales lineaires aux systemes d’equations differentielles non
  lineaires.}}  \jt{Acta. Math.}  \bvol{59},  \pg{63--87}.

\bibitem[Chandler \& Kerswell(2013)]{Chandler2013}
{\sc \au{Chandler, G.~J.} \& \au{Kerswell, R.~R.}} \yr{2013}  \at{Invariant
  recurrent solutions embedded in a turbulent two-dimensional kolmogorov flow}.
   \jt{Journal of Fluid Mechanics}  \bvol{722},  \pg{554–595}.

\bibitem[Chantry {\em et~al.\/}(2014)Chantry, Willis \& Kerswell]{Chantry2014}
{\sc \au{Chantry, M}, \au{Willis, A.~P.} \& \au{Kerswell, R.~R.}} \yr{2014}
  \at{{Genesis of streamwise-localised solutions from globally periodic
  traveling waves in pipe flow }}.  \jt{Physical Review Letters}  \bvol{112},
  \pg{164501}.

\bibitem[Cvitanovic \& Gibson(2010)]{Cvitanovic2010}
{\sc \au{Cvitanovic, P.} \& \au{Gibson, J.~F.}} \yr{2010}  \at{{ Geometry of
  the turbulence in wall-bounded shear flows: periodic orbits }}.  \jt{Physica
  Scripta}  \bvol{T142},  \pg{014007}.

\bibitem[Deguchi(2017)]{Deguchi2017}
{\sc \au{Deguchi, K.}} \yr{2017}  \at{{ Scaling of small vortices in stably
  stratified shear flows }}.  \jt{Journal of Fluid Mechanics}  \bvol{821},
  \pg{582--594}.

\bibitem[Eaves {\em et~al.\/}(2016)Eaves, Caulfield \& Mezic]{Eaves2016}
{\sc \au{Eaves, T.~S.}, \au{Caulfield, C.~P.} \& \au{Mezic, I.}} \yr{2016}
  \at{{ Transition to Turbulence: highway through the edge of chaos is charted
  by Koopman modes}}.  \jt{APS Bulletin}
  \bvol{http://meetings.aps.org/link/BAPS.2016.DFD.D8.3}.

\bibitem[Eckhardt {\em et~al.\/}(2007)Eckhardt, Schneider, Hof \&
  Westerweel]{Eckhardt2007}
{\sc \au{Eckhardt, B.}, \au{Schneider, T.~M.}, \au{Hof, B.} \& \au{Westerweel,
  J.}} \yr{2007}  \at{{ Turbulence transition in pipe flow }}.  \jt{Annual
  Review of Fluid Mechanics}  \bvol{39},  \pg{447--468}.

\bibitem[Faisst \& Eckhardt(2003)]{Faisst2003}
{\sc \au{Faisst, H.} \& \au{Eckhardt, B.}} \yr{2003}  \at{{ Traveling waves in
  pipe flow }}.  \jt{Physical Review Letters}  \bvol{91},  \pg{224502}.

\bibitem[Gibson \& Brand(2014)]{GibsonBrand2014}
{\sc \au{Gibson, J.~F.} \& \au{Brand, E.}} \yr{2014}  \at{{ Spanwise-localized
  solutions of planar shear flows}}.  \jt{Journal of Fluid Mechanics}
  \bvol{745},  \pg{25--61}.

\bibitem[Gibson {\em et~al.\/}(2008)Gibson, Halcrow \& Cvitanovic]{Gibson2008}
{\sc \au{Gibson, J.~F.}, \au{Halcrow, J.} \& \au{Cvitanovic, P.}} \yr{2008}
  \at{Visualizing the geometry of state space in plane couette flow}.
  \jt{Journal of Fluid Mechanics}  \bvol{611},  \pg{107–130}.

\bibitem[Gibson {\em et~al.\/}(2009)Gibson, Halcrow \& Cvitanovic]{Gibson2009}
{\sc \au{Gibson, J.~F.}, \au{Halcrow, J.} \& \au{Cvitanovic, P.}} \yr{2009}
  \at{Equilibrium and travelling-wave solutions of plane couette flow}.
  \jt{Journal of Fluid Mechanics}  \bvol{638},  \pg{243–266}.

\bibitem[Hall \& Sherwin(2010)]{Hall2010}
{\sc \au{Hall, P.} \& \au{Sherwin, S.}} \yr{2010}  \at{Streamwise vortices in
  shear flows: harbingers of transition and the skeleton of coherent
  structures}.  \jt{Journal of Fluid Mechanics}  \bvol{661},  \pg{178–205}.

\bibitem[Hamilton {\em et~al.\/}(1995)Hamilton, Kim \& Waleffe]{Hamilton1995}
{\sc \au{Hamilton, J.~M.}, \au{Kim, J.} \& \au{Waleffe, F.}} \yr{1995}
  \at{Regeneration mechanisms of near-wall turbulence structures}.  \jt{Journal
  of Fluid Mechanics}  \bvol{287},  \pg{317–348}.

\bibitem[Jovanovi\'c {\em et~al.\/}(2014)Jovanovi\'c, Schmid \&
  Nichols]{Jovanovic2014}
{\sc \au{Jovanovi\'c, M.~R.}, \au{Schmid, P.~J.} \& \au{Nichols, J.~W.}}
  \yr{2014}  \at{{Sparsity-promoting dynamic mode decomposition}}.  \jt{Phys.
  Fluids}  \bvol{26},  \pg{024103}.

\bibitem[Kawahara \& Kida(2001)]{Kawahara2001}
{\sc \au{Kawahara, G.} \& \au{Kida, S.}} \yr{2001}  \at{Periodic motion
  embedded in plane couette turbulence: regeneration cycle and burst}.
  \jt{Journal of Fluid Mechanics}  \bvol{449},  \pg{291–300}.

\bibitem[Kawahara {\em et~al.\/}(2012)Kawahara, Uhlmann \& van
  Veen]{Kawahara2012}
{\sc \au{Kawahara, G.}, \au{Uhlmann, M.} \& \au{van Veen, L.}} \yr{2012}
  \at{The significance of simple invariant solutions in turbulent flows}.
  \jt{Annual Review of Fluid Mechanics}  \bvol{44}~(1),  \pg{203--225}.

\bibitem[Kerswell(2005)]{Kerswell2005}
{\sc \au{Kerswell, R.~R.}} \yr{2005}  \at{{ Recent progress in understanding
  the transition to turbulence in a pipe}}.  \jt{Nonlinearity}  \bvol{18},
  \pg{R17--R44}.

\bibitem[Koopman(1931)]{Koopman1931}
{\sc \au{Koopman, B.~O.}} \yr{1931}  \at{{Hamiltonian Systems and
  Transformations in Hilbert Space}}.  \jt{Proc. Nat. Acad. Sci.}
  \bvol{17}~(5),  \pg{315--318}.

\bibitem[Kutz {\em et~al.\/}(2016{\natexlab{{\em a\/}}})Kutz, Brunton, Brunton
  \& Proctor]{DMDkutz}
{\sc \au{Kutz, J.~N.}, \au{Brunton, S.~L.}, \au{Brunton, B.~W.} \& \au{Proctor,
  J.~L.}} \yr{2016{\natexlab{{\em a\/}}}} {\em {Dynamic Mode Decomposition:
  Data-Driven Modeling of Complex Systems}\/}, 1st edn.  \publ{SIAM}.

\bibitem[Kutz {\em et~al.\/}(2016{\natexlab{{\em b\/}}})Kutz, Fu \&
  Brunton]{Kutz2016_vid}
{\sc \au{Kutz, J.~N.}, \au{Fu, X.} \& \au{Brunton, S.~L.}}
  \yr{2016{\natexlab{{\em b\/}}}}  \at{Multiresolution dynamic mode
  decomposition}.  \jt{SIAM Journal on Applied Dynamical Systems}  \bvol{15},
  \pg{713--735}.

\bibitem[Lucas {\em et~al.\/}(2017)Lucas, Caulfield \& Kerswell]{Lucas2017}
{\sc \au{Lucas, D.}, \au{Caulfield, C.~P.} \& \au{Kerswell, R.~R.}} \yr{2017}
  \at{Layer formation in horizontally forced stratified turbulence: connecting
  exact coherent structures to linear instabilities}.  \jt{Journal of Fluid
  Mechanics}  \bvol{832},  \pg{409–437}.

\bibitem[Lusch {\em et~al.\/}(2018)Lusch, Kutz \& Brunton]{Lusch2016}
{\sc \au{Lusch, B.}, \au{Kutz, J.~N.} \& \au{Brunton, S.~L.}} \yr{2018}
  \at{{Deep learning for universal linear embeddings of nonlinear dynamics}}.
  \jt{arXiv 1712.09707} .

\bibitem[Mezi\'c(2005)]{Mezic2005}
{\sc \au{Mezi\'c, I.}} \yr{2005}  \at{{Spectral Properties of Dynamical
  Systems, Model Reduction and Decompositions}}.  \jt{Nonlinear Dynam.}
  \bvol{41},  \pg{309--325}.

\bibitem[Mezi\'c(2013)]{Mezic2013}
{\sc \au{Mezi\'c, I.}} \yr{2013}  \at{{Analysis of Fluid Flows via Spectral
  Properties of the Koopman Operator}}.  \jt{Ann. Rev. Fluid Mech.}  \bvol{45},
   \pg{357--378}.

\bibitem[Nagata(1990)]{Nagata1990}
{\sc \au{Nagata, M.}} \yr{1990}  \at{Three-dimensional finite-amplitude
  solutions in plane couette flow: bifurcation from infinity}.  \jt{Journal of
  Fluid Mechanics}  \bvol{217},  \pg{519–527}.

\bibitem[Olvera \& Kerswell(2017)]{Olvera2017}
{\sc \au{Olvera, D.} \& \au{Kerswell, R.~R.}} \yr{2017}  \at{Exact coherent
  structures in stably stratified plane couette flow}.  \jt{Journal of Fluid
  Mechanics}  \bvol{826},  \pg{583–614}.

\bibitem[Page \& Kerswell(2018)]{Page2018}
{\sc \au{Page, J.} \& \au{Kerswell, R.~R.}} \yr{2018}  \at{{Koopman analysis of
  Burgers equation}}.  \jt{Physical Review Fluids}  \bvol{3},  \pg{071901(R)}.

\bibitem[Rowley \& Dawson(2017)]{Rowley2017}
{\sc \au{Rowley, C.~W.} \& \au{Dawson, S. T.~M.}} \yr{2017}  \at{{Model
  Reduction for Flow Analysis and Control}}.  \jt{Ann. Rev. Fluid Mech.}
  \bvol{49},  \pg{387--417}.

\bibitem[Rowley {\em et~al.\/}(2009)Rowley, Mezi\'c, Bagheri, Schlatter \&
  Henningson]{Rowley2009}
{\sc \au{Rowley, C.~W.}, \au{Mezi\'c, I.}, \au{Bagheri, S.}, \au{Schlatter, P.}
  \& \au{Henningson, D.~S.}} \yr{2009}  \at{{Spectral analysis of nonlinear
  flows}}.  \jt{J. Fluid Mech.}  \bvol{641},  \pg{115--127}.

\bibitem[Schmid(2010)]{Schmid2010}
{\sc \au{Schmid, P.~J.}} \yr{2010}  \at{{Dynamic mode decomposition of
  numerical and experimental data}}.  \jt{J. Fluid Mech.}  \bvol{656},
  \pg{5--28}.

\bibitem[Schneider {\em et~al.\/}(2010)Schneider, Gibson \&
  Burke]{Schneider2010}
{\sc \au{Schneider, T.~M.}, \au{Gibson, J.~F.} \& \au{Burke, J.}} \yr{2010}
  \at{Snakes and ladders: Localized solutions of plane couette flow}.
  \jt{Physical Review Letters}  \bvol{104},  \pg{104501}.

\bibitem[Schneider {\em et~al.\/}(2008)Schneider, Gibson, Lagha, De~Lillo \&
  Eckhardt]{Schneider2008}
{\sc \au{Schneider, T.~M.}, \au{Gibson, J.~F.}, \au{Lagha, M.}, \au{De~Lillo,
  F.} \& \au{Eckhardt, B.}} \yr{2008}  \at{Laminar-turbulent boundary in plane
  couette flow}.  \jt{Physical Review E}  \bvol{78},  \pg{037301}.

\bibitem[Sharma {\em et~al.\/}(2016)Sharma, Mezi\'c \& McKeon]{Sharma2016}
{\sc \au{Sharma, A.~S.}, \au{Mezi\'c, I.} \& \au{McKeon, B.~J.}} \yr{2016}
  \at{{Correspondence between Koopman mode decompositions, resolvent mode
  decomposition and invariant solutions of the Navier-Stokes equations}}.
  \jt{Phys. Rev. Fluids}  \bvol{1},  \pg{032402(R)}.

\bibitem[Tu {\em et~al.\/}(2014)Tu, Rowley, Luchtenburg, Brunton \&
  Kutz]{Tu2014}
{\sc \au{Tu, J.~H}, \au{Rowley, C.~W.}, \au{Luchtenburg, D.~M.}, \au{Brunton,
  S.~L.} \& \au{Kutz, J.~N.}} \yr{2014}  \at{{On dynamic mode decomposition:
  theory and applications}}.  \jt{J. Comput. Dynam.}  \bvol{1}~(2),
  \pg{391--421}.

\bibitem[Uhlmann {\em et~al.\/}(2010)Uhlmann, Kawahara \& Pinelli]{Uhlmann2010}
{\sc \au{Uhlmann, M.}, \au{Kawahara, G.} \& \au{Pinelli, A.}} \yr{2010}
  \at{Traveling-waves consistent with turbulence-driven secondary flow in a
  square duct}.  \jt{Physics of Fluids}  \bvol{22}~(8),  \pg{084102}.

\bibitem[Viswanath(2007)]{Viswanath2007}
{\sc \au{Viswanath, D.}} \yr{2007}  \at{Recurrent motions within plane couette
  turbulence}.  \jt{Journal of Fluid Mechanics}  \bvol{580},  \pg{339–358}.

\bibitem[Waleffe(1997)]{Waleffe1997}
{\sc \au{Waleffe, F.}} \yr{1997}  \at{{ On a self-sustaining process in shear
  flows}}.  \jt{Physics of Fluids}  \bvol{9},  \pg{883--900}.

\bibitem[Waleffe(2001)]{Waleffe2001}
{\sc \au{Waleffe, F.}} \yr{2001}  \at{Exact coherent structures in channel
  flow}.  \jt{Journal of Fluid Mechanics}  \bvol{435},  \pg{93–102}.

\bibitem[Wang {\em et~al.\/}(2007)Wang, Gibson \& Waleffe]{Wang2007}
{\sc \au{Wang, J.}, \au{Gibson, J.} \& \au{Waleffe, F.}} \yr{2007}  \at{Lower
  branch coherent states in shear flows: Transition and control}.  \jt{Physical
  Review Letters}  \bvol{98},  \pg{204501}.

\bibitem[Wedin \& Kerswell(2004)]{Wedin2004}
{\sc \au{Wedin, H.} \& \au{Kerswell, R.R.}} \yr{2004}  \at{{ Exact coherent
  structures in pipe flow: travelling wave solutions}}.  \jt{Journal of Fluid
  Mechanics}  \bvol{508},  \pg{333--371}.

\bibitem[Williams {\em et~al.\/}(2015)Williams, Kevrekidis \&
  Rowley]{Williams2015}
{\sc \au{Williams, M.~O.}, \au{Kevrekidis, I.~G.} \& \au{Rowley, C.~W.}}
  \yr{2015}  \at{{A Data-Driven Approximation of the Koopman Operator:
  Extending Dynamic Mode Decomposition}}.  \jt{J. Nonlinear Sci.}
  \bvol{25}~(6),  \pg{1307--1346}.

\bibitem[Zammert \& Eckhardt(2014)]{Zammert2014}
{\sc \au{Zammert, S.} \& \au{Eckhardt, B.}} \yr{2014}  \at{{ Streamwise and
  doubly-localized periodic orbits in plane Poiseuille flow}}.  \jt{Journal of
  Fluid Mechanics}  \bvol{761},  \pg{348--359}.

\end{thebibliography}

\end{document}